\documentclass[bibyear]{aa}  
\usepackage{natbib}
\usepackage{lastpage}
\usepackage{graphicx}
\usepackage{caption}
\usepackage{hyperref}
\usepackage{enumitem}
\usepackage{media9}
\usepackage{multirow}
\usepackage{etoolbox}
\robustify{\citet}
\hypersetup{
    colorlinks=true,
    linkcolor=blue,
    filecolor=magenta,
    urlcolor=blue,
    citecolor=blue,
    pdftitle={Overleaf Example},
    pdfpagemode=FullScreen,
    }
\usepackage{booktabs}
\usepackage{longtable}
\usepackage{txfonts}
\newcommand{\HI}{H\,{\sc i}}
\newcommand{\sigint}{\sigma_{\mathrm{int}}}
\providecommand{\nolinenumbers}{}
\begin{document}

\providecommand{\HCGSizeFractionPercent}{77}
\providecommand{\HCGSizeReductionPercent}{23}
\providecommand{\HCGDeficitPercent}{23}
\providecommand{\HCGBelowBaselinePercent}{75}
\providecommand{\HCGBelowOneSigmaPercent}{35}
\providecommand{\HCGMeanResidualDex}{-0.111}
\providecommand{\CliffProbAmigaGreaterHCGPercent}{68}
\providecommand{\HCGSampleSize}{51}
\providecommand{\HCGBelowBaselineCount}{38}
\providecommand{\AmigaSigmaDex}{0.153}
\providecommand{\PhaseOneMedianSizePercent}{77}
\providecommand{\PhaseOneReductionPercent}{23}
\providecommand{\PhaseOneDeficitPercent}{23}
\providecommand{\PhaseOneN}{24}
\providecommand{\PhaseOneMedianDeltaDex}{-0.116}
\providecommand{\PhaseTwoMedianSizePercent}{80}
\providecommand{\PhaseTwoReductionPercent}{20}
\providecommand{\PhaseTwoDeficitPercent}{20}
\providecommand{\PhaseTwoN}{20}
\providecommand{\PhaseTwoMedianDeltaDex}{-0.099}
\providecommand{\PhaseThreeCMedianSizePercent}{102}
\providecommand{\PhaseThreeCReductionPercent}{-2}
\providecommand{\PhaseThreeCDeficitPercent}{-2}
\providecommand{\PhaseThreeCN}{6}
\providecommand{\PhaseThreeCMedianDeltaDex}{+0.011}
\providecommand{\HCGSampleSizeCensored}{124}
\providecommand{\HCGUpperLimitCount}{73}
\providecommand{\HCGKMMedianDex}{-0.535}
\providecommand{\HCGKMMedianBound}{1}
\providecommand{\HCGKMDeficitPercent}{71}
\providecommand{\HCGKMSizeFractionPercent}{29}
\providecommand{\HCGKMBelowBaselinePercent}{90}
\providecommand{\CliffProbAmigaGreaterHCGCensoredPercent}{83}
\providecommand{\CliffDeltaCensored}{0.70}
\providecommand{\GehanZ}{-11.9}
\providecommand{\GehanP}{1.1\times10^{-32}}
\providecommand{\PhaseOneNCensored}{36}
\providecommand{\PhaseOneNLim}{12}
\providecommand{\PhaseOnePctBelow}{86}
\providecommand{\PhaseTwoNCensored}{33}
\providecommand{\PhaseTwoNLim}{13}
\providecommand{\PhaseTwoPctBelow}{85}
\providecommand{\PhaseThreeCNCensored}{37}
\providecommand{\PhaseThreeCNLim}{31}
\providecommand{\PhaseThreeCPctBelow}{92}
\providecommand{\PhaseThreeANCensored}{18}
\providecommand{\PhaseThreeANLim}{17}
\providecommand{\PhaseThreeAPctBelow}{100}
\providecommand{\AmigaSlope}{0.455}
\providecommand{\AmigaIntercept}{-2.732}
\providecommand{\AmigaScatterDex}{0.080}
\providecommand{\AmigaN}{35}
\providecommand{\HcgsSlope}{0.509}
\providecommand{\HcgsIntercept}{-3.265}
\providecommand{\HcgsScatterDex}{0.077}
\providecommand{\HcgsN}{53}
\providecommand{\AmigaHcgsSlope}{0.496}
\providecommand{\AmigaHcgsIntercept}{-3.144}
\providecommand{\AmigaHcgsScatterDex}{0.078}
\providecommand{\AmigaHcgsN}{88}
\providecommand{\MighteeSlope}{0.503}
\providecommand{\MighteeIntercept}{-3.268}
\providecommand{\MighteeScatterDex}{0.053}
\providecommand{\MighteeN}{204}
\providecommand{\WangSixteenSlope}{0.506}
\providecommand{\WangSixteenIntercept}{-3.291}
\providecommand{\WangSixteenScatterDex}{0.063}
\providecommand{\WangSixteenN}{435}
\providecommand{\MighteeWangSlope}{0.504}
\providecommand{\MighteeWangIntercept}{-3.276}
\providecommand{\MighteeWangScatterDex}{0.060}
\providecommand{\MighteeWangN}{639}
\providecommand{\CombinedFitSlope}{0.508}
\providecommand{\CombinedFitIntercept}{-3.305}
\providecommand{\CombinedFitScatterDex}{0.065}
\providecommand{\CombinedFitN}{727}
\providecommand{\AmigaSlopeTension}{1.37}
\providecommand{\AmigaInterceptTension}{1.57}
\providecommand{\HcgsSlopeTension}{0.18}
\providecommand{\HcgsInterceptTension}{0.05}

   \title{The size of the HI disk across different environments: isolated, compact groups, clusters, and pairs}
   \author{R. Ianjamasimanana\inst{1}
          \and L. Verdes-Montenegro\inst{1}
          \and K. M. Hess\inst{2,3}
          \and P. Kamphuis\inst{1}
          \and M. G. Jones\inst{4}
          \and J. Garrido\inst{1}
          \and S. H. A. Rajohnson\inst{5}
          \and A. Sorgho\inst{1}
          \and B. Namumba\inst{1}
          \and S. Sanchez-Exp\'osito\inst{1}
          \and M. Korsaga\inst{1}
          }

   \institute{Instituto de Astrof\'isica de Andaluc\'ia (CSIC), Glorieta de la Astronom\'ia s/n, 18008 Granada, Spain\\
              \email{ianja@iaa.es}
         \and Department of Space, Earth and Environment, Chalmers University of Technology, Onsala Space Observatory, 43992 Onsala, Sweden
         \and ASTRON, the Netherlands Institute for Radio Astronomy, Oude Hoogeveensedijk 4, 7991 PD Dwingeloo, The Netherlands
         \and IPAC, Mail Code 100-22, Caltech, 1200 E. California Blvd., Pasadena, CA 91125, USA
         \and INAF -- Osservatorio Astronomico di Cagliari, Via della Scienza 5, I-09047 Selargius (CA), Italy
             }

   \date{Received June 28, 2026; accepted July 15, 2026}

\abstract
{The 21 cm line of atomic hydrogen (\HI)\ is a sensitive tracer of the outer disk of galaxies, where environmental signatures are most apparent. The relative extent of \HI\ disks compared to optical disks ($\mathrm{D_{HI}~vs~D_{25}}$) is thought to provide a quantitative measure of such imprint, yet systematic comparisons between extreme environments remain scarce.}
{We quantify the relative extent of \HI\ disks compared to stellar disks in Hickson Compact Groups (HCGs) and in the Analysis of the interstellar Medium in Isolated GAlaxies (AMIGA) sample. We aim to establish how the compact group environment shapes the extent of the \HI\ disk by using AMIGA as a control sample that captures secular evolution with minimal external influence.}
{We calculate the \HI\ diameters of AMIGA and HCG samples by directly fitting an ellipse to the 1~$\mathrm{M_{\odot}\,pc^{-2}}$ iso-density contour. Because $D_{\rm HI}$ and $D_{25}$ are nonlinearly related, we avoid the traditional $D_{\rm HI}/D_{25}$ ratio, which carries a size-dependent bias, and instead quantify truncation as the residual from the isolated-galaxy $D_{\rm HI}$--$D_{25}$ baseline. We establish the \(D_{\rm HI}\)-\(D_{25}\) scaling relation for AMIGA via Bayesian analysis. We then quantify how HCGs deviate from this baseline. To maximize the reproducibility of this analysis, we provide a fully reproducible framework via an installable Python package and a Snakemake workflow that reproduce all analysis, scientific products, and generate the manuscript.}
{HCG galaxies lie systematically below the isolated galaxy baseline in the $D_{\mathrm{HI}}$--$D_{25}$ plane. When members with \HI\ nondetections are included as upper limits, HCGs have \HI\ disks that are at least ${\sim}71\%$ smaller than expected for isolated galaxies of the same optical diameter. The \HI\ disk truncation depends strongly on HCG evolutionary sequence, increasing monotonically from Phase~1 to Phase~3. A comparison with literature samples places HCGs at the most-truncated end, statistically indistinguishable from the Virgo cluster sample (VIVA).}
{Compared to AMIGA, \HI\ disks are typically smaller relative to the optical disk in loose groups, compact groups, and cluster infall/field environments, and are most strongly truncated in HCGs and in the Virgo cluster sample.}

   \keywords{galaxies:evolution --
                galaxies:groups --
                galaxies:interactions --
                galaxies: ISM
               } 
   \maketitle
   \nolinenumbers

\section{Introduction}
The evolution of galaxies is significantly influenced by their surroundings. Galaxies living in dense environments, such as galaxy clusters or groups, are known to experience interactions and mergers that significantly alter their structure and gas content \citep{1984ApJ...281...95P, 1997ApJ...490..577D}. Conversely, galaxies in isolated environments are thought to evolve mostly through internal processes, making them ideal laboratories to study intrinsic galaxy properties with minimal external influences \citep{2013A&A...560A...9A, 2018A&A...609A..17J}.\\ \indent The extent of a galaxy's atomic hydrogen (\HI) disk, measured here by the diameter $D_{\mathrm{HI}}$ at which the \HI\ surface density reaches $1~\mathrm{M_{\odot}\,pc^{-2}}$, relative to its optical disk, often measured by the isophotal diameter $D_{\mathrm{25}}$ at $25$ $B$-mag arcsec$^{-2}$, has been known to be a sensitive tracer of environmental influences \citep[e.g.,][]{1988A&AS...72..427W, 1994AJ....107.1003C, 2009AJ....138.1741C, 2022MNRAS.510.1716R} as opposed to the well-known size--mass relation. The latter remains tight across different environments \citep{2016MNRAS.460.2143W}. The $\mathrm{D_{HI}-D_{25}}$ relation, in contrast, captures the differential response of gas and stars to external perturbations. This is because, while the stellar component traces the long-term mass assembly of a galaxy, the atomic hydrogen extends to much larger radii \citep[e.g.,][]{1981AJ.....86.1791B}  where the gravitational restoring force is weaker, making it vulnerable to gas removal mechanisms like ram-pressure stripping, tidal interactions, and starvation \citep{1984ApJ...281...95P, 1997ApJ...490..577D}. The strength of these processes is thought to depend on the environment, but quantifying their imprint on the \HI\ disk requires a well-defined reference sample. Compact groups are among the densest local environments, and isolated galaxies have evolved with minimal external influence. They define the opposite ends of the local environmental density spectrum, motivating the central question of this paper: how strongly does the compact-group environment truncate the \HI\ disk relative to an equivalent isolated galaxy of the same optical size? \\ \indent Previous work used the $D_{\mathrm{HI}}/D_{25}$ ratio as a quantitative measure of the environmental effect on the size of the H\,{\sc i} disk relative to the stellar disk, without explicitly testing whether this ratio provides an unbiased metric when $D_{\mathrm{HI}}$ and $D_{25}$ are nonlinearly related \citep{B1997, 2005A&A...442..137N, 2009AJ....138.1741C, 2016MNRAS.460.2143W}.
\\ \indent \citet{1984AJ.....89..758H} established empirical scaling relations between a galaxy's \HI\ mass and its optical diameter, blue luminosity, and morphological type, using a sample of relatively isolated, unperturbed galaxies taken to define the standards of \HI\ normalcy. These relations provide the reference framework for predicting the expected \HI\ mass of a galaxy and hence for quantifying its \HI\ deficiency. Applying this framework, \citet{2001ApJ...548...97S} found that spiral galaxies in nearby clusters are systematically more \HI-deficient than field galaxies of the same type and size, with the deficiency increasing toward the cluster centers and remaining detectable out to roughly two Abell radii. They concluded that hydrodynamic interactions with the intracluster medium, in particular ram-pressure stripping, are the most plausible cause of this gas removal. Comprehensive summaries of this and related observational evidence for gas stripping in dense environments are given in the recent reviews of \citet{2021PASA...38...35C} and \citet{2022A&ARv..30....3B}. Beyond clusters, \HI\ deficiency has also been reported in loose groups \citep[e.g.,][]{2013MNRAS.436...34C}, in compact groups \citep{2001A&A...377..812V, 2004PASA...21..318S, 2015ApJ...812...78B, 2023A&A...670A..21J}, and in interacting pairs \citep{1996AJ....111..655H, 2015MNRAS.449.3719S}. Given the tightness of the \HI\ size--mass relation, this \HI\ deficiency is expected to translate into a corresponding reduction of the \HI\ disk extent, as indeed observed in nearby clusters such as Virgo \citep{1990AJ....100..604C} and Coma \citep{2000AJ....119..580B}. In cluster galaxies, truncation is not limited to the atomic gas; it has also been observed in the other phases of the interstellar medium and in the young stellar population \citep[e.g.,][]{2004ApJ...613..866K, 2022A&ARv..30....3B}, making gas truncation a key ingredient of the quenching of star formation in rich environments \citep{2006PASP..118..517B}. Because ram pressure is a hydrodynamic process, it acts on the gas but not on the stellar component \citep{2022A&ARv..30....3B}; it removes the loosely bound, low-density gas of the outer \HI\ disk first, leaving the stellar disk largely unaffected. \citet{B1997} compiled $\mathrm{D_{HI}/D_{25}}$ measurements for 108 spiral and irregular galaxies and found an average value of $\sim$1.7. \citet{2017ASSL..434..209B} compiled data from 16 surveys and showed that the peak \HI-to-optical size ratio varies considerably between samples. It ranges from $\sim$1.0 for cluster samples to $\sim$1.7 for field spirals, with only $\sim$10\% of galaxies exceeding $\mathrm{D_{HI}/D_{25}} \sim 3$. The VLA Imaging of Virgo Spirals in Atomic Gas (VIVA) by \citet{2009AJ....138.1741C} also showed that galaxies near the cluster core (defined by those authors as projected distances $d_{87} \lesssim 0.5$~Mpc from M87) have \HI\ disks truncated well inside the optical disk ($D_{\rm HI}/D_{25} < 0.5$), often with asymmetric \HI\ morphologies indicative of ongoing ram-pressure stripping. More recently, \citet{2022MNRAS.510.1716R} measured the ratio of \HI\ diameter ($d_{\rm HI}$, defined at the same $1~\mathrm{M_{\odot}\,pc^{-2}}$ level as our $D_{\rm HI}$) to $r$-band optical diameter ($d_{\rm opt}$, at 23.5 mag arcsec$^{-2}$) for galaxies in and around the Hydra I cluster using WALLABY observations. They found that \HI -detected cluster and infall galaxies have systematically lower \HI-to-optical diameter ratios than field galaxies, with the median ratio decreasing from $d_{\rm HI}/d_{\rm opt}=3.5$ in the field to $d_{\rm HI}/d_{\rm opt}=2.0$ in the cluster. Only a small fraction of the \HI-detected cluster galaxies have ratios below unity, while the \HI\ nondetection upper limits are consistent with $d_{\rm HI}/d_{\rm opt}<1$. The ratios tend to be smaller for galaxies closer to the cluster center than for those at larger distances. These \HI-detected galaxies still lie on the star-forming main sequence, suggesting that the gas removal traced by WALLABY mainly affects the outer gas reservoir and has not yet strongly affected the inner star-forming disk. The interplay between the \HI\ content of galaxies in rich environments and their position on the main sequence has been characterized extensively in the literature \citep[see][and references therein]{2023A&A...669A..73B}. \\ \indent While previous studies have focused primarily on cluster environments or heterogeneous field samples, comparisons between galaxies at opposite extremes of the environmental spectrum have not been done, yet such comparisons would provide the clearest test of how environment affects \HI\ disk extent. The Analysis of the interstellar Medium in Isolated GAlaxies project \citep[AMIGA;][]{2005A&A...436..443V} provides a sample at one such extreme. 
AMIGA has used the strictest isolation criteria ever applied to galaxies based on the local number density of neighboring galaxies ($\eta_{k}$) and the tidal strength ($Q_k$) to select a sample of 791 isolated galaxies \citep{2013A&A...560A...9A}. These galaxies are, in principle, virtually unaffected by interactions in the last $\sim$ 3 Gyr \citep{2007A&A...472..121V, 2018A&A...609A..17J}. They are more isolated than typical field galaxies \citep[e.g.,][]{2007A&A...462..507L} considered in major surveys. They serve as benchmarks for studying secular evolution in the most extreme case of isolation, offering a valuable point of comparison to galaxies living in denser environments. Previous studies have revealed that, compared to galaxies in denser environments, AMIGA galaxies exhibit lower scatter in several global relations and distributions, including the SFR--$M_{\star}$ relation, the gas-fraction and \HI\ deficiency distributions, and the distribution of \HI\ profile asymmetries \citep{2011A&A...532A.117E, 2018A&A...609A..17J, 2020MNRAS.499.3193B}. \citet{2013MNRAS.434..325F} found that isolated late-type AMIGA galaxies are about 1.2 times larger in stellar effective radius than less isolated galaxies of the same stellar mass. 
At the opposite extreme, Hickson Compact Groups \citep[HCGs,][]{1982ApJ...255..382H} are among the densest galaxy environments in the local Universe. These groups are small systems of four to ten galaxies, located in very close proximity to each other. Observations have shown that galaxies in HCGs exhibit disturbed morphologies \citep{2025A&A...696A.176I}, enhanced star formation \citep{2010ApJ...716..556T}, depleted \HI\ content \citep{2025A&A...696A.177S}, and truncated gas disks \citep{2023A&A...670A..21J}, indicating strong environmental processing. \\ \indent In this paper, we quantify how much the compact-group environment truncates the \HI\ disk by comparing HCGs against AMIGA as the isolated-galaxy reference. Specifically, we (a) establish the AMIGA $D_{\mathrm{HI}}$--$D_{25}$ relation as a baseline, and (b) measure how HCGs and a range of literature samples spanning clusters, loose groups, and pairs deviate from this reference. The paper is organized as follows. Section~\ref{sample} describes the AMIGA and HCG samples. Section~\ref{methodology} presents the measurement of $D_{\rm HI}$ from the $1~M_\odot\,{\rm pc^{-2}}$ iso-density contour, the fitting procedure to define the AMIGA $D_{\rm HI}$--$D_{25}$ baseline, the residual-based method to quantify \HI\ disk truncation, and a test of whether AMIGA and HCGs follow the established \HI\ size--mass relation. Section~\ref{results} compares how galaxies in different environments deviate from the AMIGA $D_{\rm HI}$--$D_{25}$ isolated-galaxy reference. Section~\ref{discussion} discusses what these deviations imply for \HI\ disk truncation in different environments. Finally, Sect.~\ref{summary} summarizes the main conclusions of the paper. Apart from the scientific results presented here, we have developed a full analysis framework following the FAIR principles (Findable, Accessible, Interoperable, and Reusable) to ensure reproducibility and promote open science. The analysis is distributed as an installable software package and an automated workflow that regenerate all quantitative results and the manuscript itself from a single set of publicly available inputs. We describe this framework and the tools required to reproduce our work in its entirety in Appendix~\ref{reproducibility}.
\section{Sample}\label{sample}
\subsection{Isolated galaxies from the AMIGA project}
Our sample of isolated galaxies comes from the AMIGA project. The sample was originally built on the Catalogue of Isolated Galaxies \citep[CIG;][]{1973SoSAO...8....3K}, whose 1051 members were selected such that no comparable-size neighbor (within a factor of four in diameter) lies within 20 diameters of the candidate galaxy. AMIGA refined this catalog with improved positions, optical photometry, morphologies, and radial velocities, and requantified the isolation of each galaxy through two complementary parameters, the local number density of neighbors ($\eta_{k}$) and the tidal strength exerted on the primary ($Q_{k}$), estimated from a systematic search for neighbors in digitized sky survey images \citep{2007A&A...472..121V, 2007A&A...470..505V, 2013A&A...560A...9A}. Galaxies satisfying both criteria form the final sample of 791 strictly isolated galaxies. They represent the extremely low-density end of the galaxy population in the nearby Universe. 
The AMIGA selection criteria ensure that the catalog is free from nearby, similarly sized companions that likely exert significant gravitational influence on the target galaxies and the effects of past interactions have likely been washed out \citep{2007A&A...472..121V}. 
For a detailed discussion of how the AMIGA sample was selected, we refer the reader to \citet{2007A&A...472..121V, 2007A&A...470..505V} and \citet{2013A&A...560A...9A}.
\\ \indent 
AMIGA has 35 galaxies with resolved \HI\ data, which we use to define our sample. We refer to this as our resolved AMIGA sample. We list their properties in Table~\ref{table:obs_prop}, including the \HI\ and optical diameters studied in this work. To extend our sample of isolated galaxies, we include single-dish measurements of 372 AMIGA galaxies from \citet{2018A&A...609A..17J}, after excluding 16 galaxies that overlap with our resolved sample, five that were excluded by those authors since they fall well outside the expected $M_{\mathrm{HI}}$--$L_{B}$ and $M_{\mathrm{HI}}$--$D_{25}$ scaling relations for isolated galaxies, and six galaxies with no measured $D_{25}$. We explain this more in the methodology section. The AMIGA sample lacks low-mass galaxies. The range of stellar mass considered in this analysis is $9.14 \le \log(M_{\star}/\mathrm{M_{\odot}}) \le 10.94$.  

\begin{table*}
\centering
\caption{\label{table:obs_prop}Properties of galaxies from AMIGA}
\begin{tabular}{lccccccccc}
\toprule \toprule
CIG   & Telescopes& $B_{min}$ & $B_{maj}$  & res.  & $N_{\mathrm{HI}}~(3\sigma)$ &log(M$_{\star}$) & $\mathrm{D_{25}}$ & $\mathrm{D_{HI}}$ & $Err_{\mathrm{D_{HI}}}$\\
     &	          & [\arcsec] & [\arcsec]  & [kpc] & [$10^{19}~cm^{-2}$]         &[M$_{\odot}$]    &[kpc]              & [kpc] & [kpc]\\
\midrule
96   & VLA       & 15.6  & 16.8 & 2.3  & 7.7  & 10.04  &  21.99   & 81.85 &  0.42\\
102  & WSRT      & 24.51 & 29.5 & 12.4 & 3.1  & 10.85  &  10.16  & 100.40&  1.44\\
103  & VLA       & 45.5  & 52.9 & 6.9  & 2.8  & 9.97   &  18.41   & 41.09 &  0.35\\
123  & VLA       & 17.6  & 15.7 & 8.9  & 15.0 & 10.94  &  41.30  & 101.74&  0.67\\
147  & WSRT      & 33.1  & 23.5 & 7.1  & 2.9  & 10.76  &  18.98  & 74.97 &  0.76\\
159  & WSRT      & 23.0  & 29.7 & 10.4 & 3.2  & 10.68  &  55.00  & 63.21 &  2.04\\
188  & GMRT      & 43.1  & 36.6 & 6.7  & 4.1  & 9.72   &  23.00  & 30.03 &  0.17\\
232  & WSRT      & 32.9  & 19.6 & 12.8 & 3.1  & 10.23  &  33.36  & 50.41 &  0.87\\
240  & VLA       & 54.4  & 62.1 & 25.8 & 1.7  & 10.56  &  26.44  & 37.12 &  0.94\\
292  & VLA       & 46.6  & 42.1 & 7.1  & 1.6  & 10.14  &  20.09  & 39.68 &  0.13\\
314  & WSRT      & 28.0  & 27.7 & 6.5  & 3.2  & 10.52  &  21.17  & 52.84 &  0.52\\
329  & VLA       & 13.9  & 13.5 & 4.9  & 10.6 & 10.88  &  39.83  & 81.07 &  0.97\\
359  & VLA       & 65.9  & 60.8 & 26.8 & 0.8  & 10.76  &  19.63  & 24.02 &  0.67\\
361  & VLA       & 16.3  & 13.5 & 9.4  & 12.1 & 10.94  &  42.19  & 78.75 &  0.69\\
421  & VLA       & 56.3  & 50.0 & 32.7 & 1.4  & 10.66  &  36.01  & 75.68 &  0.46\\
463  & VLA       & 45.6  & 42.5 & 8.5  & 1.4  & 9.70   &  20.57  & 33.41 &  0.16\\
512  & GMRT      & 13.3  & 17.3 & 2.1  & 22.7 & 9.75   &  12.53   & 18.40 &  0.10\\
551  & VLA       & 47.6  & 43.1 & 9.6  & 1.2  & 9.24   &  17.96  & 42.27 &  0.21\\
553  & Apertif   & 25.3  & 12.8 & 12.6 & 23.2 & 10.83  &  37.75  & 57.94 &  1.16\\
571  & VLA       & 52.3  & 67.2 & 13.7 & 5.5  & 9.70   &  10.57  & 35.92 &  0.26\\
581  & Apertif   & 30.9  & 14.6 & 14.2 & 18.2 & 10.88  &  7.20  & 96.92 &  1.13\\
587  & Apertif   & 25.1  & 17.9 & 15.1 & 21.0 & ---    &  24.95  & 45.49 &  1.44\\
604  & WSRT      & 33.2  & 28.7 & 4.9  & 0.7  & 10.6   &  24.22  & 40.63 &  0.15\\
626  & GMRT      & 30.0  & 30.0 & 3.7  & 9.3  & 10.05  &  16.42   & 19.73 &  0.14\\
660  & VLA       & 58.8  & 46.3 & 11.0 & 2.3  & 9.14   &  9.03   & 24.92 &  0.17\\
676  & Apertif   & 17.3  & 13.2 & 8.2  & 45.2 & 10.62  &  46.43  & 61.77 &  1.44\\
736  & VLA       & 58.0  & 52.2 & 7.2  & 1.9  & 10.38  &  24.67   & 25.66 &  0.26\\
744  & VLA       & 67.2  & 55.7 & 14.9 & 0.7  & 9.32   &  19.54  & 37.92 &  0.19\\
812  & VLA       & 53.7  & 46.5 & 14.5 & 0.9  & ---    &  14.71  & 93.80 &  0.66\\
983  & Apertif   & 24.5  & 13.7 & 9.0  & 22.4 & 10.58  &  37.49  & 52.45 &  0.77\\
988  & Apertif   & 29.7  & 13.4 & 15.6 & 23.4 & 10.83  &  57.30  & 65.46 &  1.40\\
1000 & Apertif   & 23.5  & 13.5 & 10.0 & 24.0 & 9.83   &  26.05  & 42.12 &  1.07\\
1004 & VLA       & 130.0 & 48.9 & 21.9 & 0.6  & 10.52  &  34.33  & 62.09 &  0.19\\
1006 & Apertif   & 25.0  & 14.1 & 10.5 & 16.1 & 10.42  &  20.44  & 31.02 &  0.61\\
1019 & VLA       & 56.4  & 47.9 & 17.2 & 1.3  & 10.34  &  21.06  & 82.28 &  0.78\\
\bottomrule
\end{tabular}
\tablefoot{Columns: (1) CIG ID number, (2) instruments used to observe the galaxies, (3--4) minor and major axis of the beam, (5) linear resolution of the observations, (6) column density sensitivity limit over 20 $\mathrm{km~s^{-1}}$, (7) stellar mass, (8) Optical diameter from the HyperLeda database, (9) \HI\ diameter, (10) Errors in \HI\ diameter.}
\end{table*}

\subsection{Hickson Compact Groups sample}
HCGs are a distinct class of galaxy groups known for their dense configurations that promote frequent interactions among their members. In total, 100 such groups were cataloged by \citet{1982ApJ...255..382H}. A reassessment by \citet{1997ApJ...482..640S} found that 82 systems satisfy Hickson's original selection criteria; among these, 61 are bona fide compact groups, defined as accordant-redshift systems containing at least four galaxies. The remaining 39 systems in the original Hickson catalog were classified as false compact groups. Galaxies in HCGs undergo extreme preprocessing of their \HI\ gas before falling into clusters. According to \citet{2001A&A...377..812V}, the \HI\ gas in HCGs follows an evolutionary progression. Initially, \HI\ is bound within the galaxy members (Phase 1). As interactions intensify, \HI\ is progressively stripped and relocated into the intragroup medium, forming tidal features (Phase 2). In the final stage (Phase 3), the member galaxies are either completely devoid of their \HI\ gas or are severely \HI-deficient. \citet{2023A&A...670A..21J} reassessed the definition of the evolutionary sequence of HCGs by redefining the \HI\ content of each phase. In their scheme, Phase 1 corresponds to groups where less than 25\% of the detected \HI\ is found in extended structures, with the majority of the gas still confined to individual galaxies. Phase 2 includes groups where between 25\% and 75\% of their \HI\ is found in extended features. Phase 3a describes groups in which more than 75\% of the \HI\ has been stripped from the galactic disks and resides in extended structures, or where the gas is no longer detectable. In addition, they removed the Phase 3b defined by \citet{2001A&A...377..812V} and introduced a new subphase, Phase 3c, representing systems where only one member retains significant \HI\@. Whether these single gas-rich members are recently accreted galaxies or survivors is not well established. If they have been recently incorporated into the groups, we expect their \HI\ disk sizes to be similar to those of galaxies in Phase 1 groups. \\ \indent 
Because gas removal and redistribution are highly efficient in HCGs, they provide a natural laboratory to measure the effects of the environments on the \HI\ disks of galaxies. We therefore draw our sample of interacting galaxies from the 38 HCGs analyzed by \citet{2023A&A...670A..21J}.  
They used archival VLA \HI\ data to visually separate features associated with the galaxies from those in the intragroup medium. After excluding four groups because of data-quality issues (HCG 38, 47, 49, and 57), two misclassified compact groups (HCG 48 and 54), and two groups (HCG 95 and HCG 100) whose members could not be reliably separated from the intragroup medium, our final sample comprises 30 HCGs. Among these, 51 members in 25 groups have measured \HI\ diameters and $D_{25}$, and their properties are listed in Table~\ref{table:obs_prop_hcgs}. To avoid biasing the truncation analysis against the most gas-poor systems, we also include the 73 \HI\ nondetected members in these groups, adopting the beam size as an upper limit on their \HI\ diameter (see Table~\ref{table:upperlimits}). In total, we have 124 HCG members; they cover the entire evolutionary sequence of HCGs, ranging from galaxies with regular morphologies to highly disturbed ones. 

\begin{table*}
\centering
\caption{\label{table:obs_prop_hcgs}Properties of galaxies from HCGs}
\begin{tabular}{lcccccccccc}
\toprule \toprule
HCG & Phase & members & $B_{\mathrm{min}}$ & $B_{\mathrm{maj}}$ & res. & $N_{\mathrm{HI}}~(3\sigma)$&\HI\--def & $D_{\mathrm{HI}}$ & $Err_{D_{\mathrm{HI}}}$ & $D_{25}$ \\
    &      &          &  [\arcsec]         & [\arcsec]          & [kpc] & [$10^{19}~cm^{-2}$]  & dex       & [kpc]       & [kpc]      & [kpc]     \\
\midrule
\multirow{3}{*}{2} & \multirow{3}{*}{1}  & HCG2a    &   \multirow{3}{*}{69.3} & \multirow{3}{*}{51.4}  &20.9&6.2& -0.34  &  50.23 & 0.32  & 21.51 \\
                   &                     & HCG2b    &                         &                        &20.9&5.1& 0.3    &  22.19 & 0.34  & 12.96 \\
                   &                     & HCG2c    &                         &                        &20.5&3.3& -0.07  &  29.97 & 0.20  & 14.59 \\
\multirow{3}{*}{7} & \multirow{3}{*}{1}  & HCG7a    &   \multirow{3}{*}{35.0} & \multirow{3}{*}{28.0}  &10.9&2.2& 0.68   &  28.54 & 0.53  & 29.70 \\
                   &                     & HCG7c    &                         &                        &11.3&2.4& 0.34   &  40.79 & 0.32  & 25.69 \\
                   &                     & HCG7d    &                         &                        &10.6&2.0& 0.16   &  18.23 & 0.29  & 13.30 \\
\multirow{2}{*}{10}& \multirow{2}{*}{1}  & HCG10a   &   \multirow{2}{*}{61.6} & \multirow{2}{*}{49.7}  &21.5&4.5& 0.06   &  79.50 & 0.58  & 48.07 \\
                   &                     & HCG10d   &                         &                        &18.8&1.0& 0.26   &  15.97 & 0.20  & 14.59 \\
15                 &        3c           & HCG15f   &  60.7                   & 47.5                   &31.4&1.7& 0.13   &  21.86 & ---   & 16.52 \\
\multirow{5}{*}{16}& \multirow{5}{*}{2}  & NGC848   &  \multirow{5}{*}{38.9}  & \multirow{5}{*}{31.7}  &11.9&3.9& -0.07  &  36.92 & 0.32  & 19.23 \\
                   &                     & HCG16a   &                         &                        &12.4&1.7& 0.82   &  25.13 & 0.20  & 12.92 \\
                   &                     & HCG16b   &                         &                        &11.9&3.5& 0.94   &  13.06 & 0.21  & 23.13 \\
                   &                     & HCG16c   &                         &                        &11.7&4.6& 0.21   &  28.72 & 0.34  & 18.84 \\
                   &                     & HCG16d   &                         &                        &11.7&8.6& -0.09  &  44.70 & 0.30  & 20.65 \\
\multirow{2}{*}{19}&  \multirow{2}{*}{1} & HCG19b   &   \multirow{2}{*}{41.0} & \multirow{2}{*}{27.0}  &12.4&1.6& -0.24  &  11.64 & 0.20  & 13.48 \\
                   &                     & HCG19c   &                         &                        &12.6&1.2& 0.6    &  24.83 & 0.16  & 18.97 \\
22                 &        3c           & HCG22c   &                    50.9 & 36.8                   &11.0&3.7& -0.27  &  35.11 & 0.44  & 14.79 \\
\multirow{4}{*}{23}& \multirow{4}{*}{1}  & HCG23-26 &   \multirow{4}{*}{25.3} & \multirow{4}{*}{20.0}  &11.1&5.8& -1.64  &  27.40 & 0.60  & ---   \\
                   &                     & HCG23a   &                         &                        &9.9&6.6&0.16   &  39.03   & 0.75  & 29.04 \\
                   &                     & HCG23b   &                         &                        &10.2&9.0&-0.34  &  43.73  & 0.70  & 31.38 \\
                   &                     & HCG23d   &                         &                        &9.2&6.0&-0.5   &  27.35   & 0.65  & 11.60 \\
\multirow{2}{*}{25}&  \multirow{2}{*}{1} & HCG25a   &   \multirow{2}{*}{65.4} & \multirow{2}{*}{56.6}  &34.0&1.0&-0.15  &  54.52  & 0.32  & 31.78 \\
                   &                     & HCG25b   &                         &                        &34.8&0.8&0.03   &  36.16  & 0.40  & 23.06 \\
\multirow{2}{*}{26}&   \multirow{2}{*}{1}& HCG26a   &   \multirow{2}{*}{25.5} & \multirow{2}{*}{16.8}  &19.6&7.3&-0.83  &  63.57  & 1.16  & 50.65 \\
                   &                     & HCG26e   &                         &                        &19.4&7.3&0.1    &  20.60  & 0.96  & 13.21 \\
\multirow{5}{*}{31}&  \multirow{5}{*}{2} & HCG31a   &   \multirow{5}{*}{14.6} & \multirow{5}{*}{12.1}  &4.7&32.6&-0.14  &  15.41  & 0.13  & 18.19 \\
                   &                     & HCG31b   &                         &                        &5.1&16.9&0.15   &  17.85  & 0.20  & 14.93 \\
                   &                     & HCG31c   &                         &                        &4.9&22.5&1.0    &  16.02  & 0.30  & 9.95 \\
                   &                     & HCG31g   &                         &                        &4.8&19.3&-0.02  &  13.85  & 0.21  & 8.12 \\
                   &                     & HCG31q   &                         &                        &4.9&17.1&-0.07  &  12.38  & 0.18  & ---   \\
33                 &        3c           & HCG33c   &                    17.5 &                  15.4  &12.1&19.3&-0.96  &  48.17 & 0.63  & 25.89 \\
\multirow{2}{*}{40}&  \multirow{2}{*}{2} & HCG40c   &   \multirow{2}{*}{58.8} & \multirow{2}{*}{44.8}  &36.9&1.0&0.18   &  28.06  & 0.54  & 46.40 \\
                   &                     & HCG40d   &                         &                        &35.1&3.1&0.61   &  24.39  & ---   & 26.00 \\
56                 &        3c           & HCG56a   &                    23.5 &                  19.6  &17.5&2.8&-0.32  &  62.54  & 0.91  & 34.32 \\
\multirow{2}{*}{58}&  \multirow{2}{*}{2} & HCG58a   &   \multirow{2}{*}{65.8} & \multirow{2}{*}{57.0}  &35.9&1.0&0.27   &  43.50  & 0.35  & 33.35 \\
                   &                     & HCG58e   &                         &                        &35.0&0.4&0.32   &  24.63  & ---   & 16.70 \\
\multirow{3}{*}{59}&  \multirow{3}{*}{1} & HCG59b   &   \multirow{3}{*}{24.0} & \multirow{3}{*}{16.0}  &8.3&5.0&-0.31  &  35.20   & 0.67  & 11.34 \\
                   &                     & HCG59c   &                         &                        &8.8&1.7&0.71   &  14.84   & 0.46  & 15.24 \\
                   &                     & HCG59d   &                         &                        &8.1&5.1&-0.34  &  28.16   & 0.42  & 10.17 \\
61                 &        3a           & HCG61c   &                   26.2  &                 18.8   &9.4&2.7&0.54   &  23.54   & 0.35  & 28.31 \\
68                 &        3c           & HCG68c   &                   58.3  &                 54.8   &13.2&0.1&-0.12  &  67.23  & 0.37  & 26.62 \\
\multirow{4}{*}{71}&  \multirow{4}{*}{2} & AGC24021 &   \multirow{4}{*}{26.0} & \multirow{4}{*}{20.0}  &20.8&2.6&-0.75  &  47.90  & 0.62  & 11.78 \\
                   &                     & AGC732898&                         &                        &20.5&2.2&-0.58  &  29.16  & 1.13  & 13.62 \\
                   &                     & HCG71a   &                         &                        &21.2&2.7&-0.16  &  11.73  & 0.97  & 41.45 \\
                   &                     & HCG71c   &                         &                        &19.6&2.8&-0.31  &  53.88  & 1.00  & 17.52 \\
79                 &        2            & HCG79d   &                    20.8 &                  17.4  &8.0&5.7&-0.48  &  22.95   & 0.52  & 19.46 \\ 
\multirow{4}{*}{88}&  \multirow{4}{*}{1} & HCG88a   &   \multirow{4}{*}{22.6} & \multirow{4}{*}{17.2}  &10.2&4.8&1.15   &  19.52  & 0.51  & 32.58 \\
                   &                     & HCG88b   &                         &                        &10.2&6.1&0.62   &  37.03  & 0.44  & 27.10 \\
                   &                     & HCG88c   &                         &                        &10.3&6.3&-0.08  &  55.13  & 0.46  & 22.32 \\
                   &                     & HCG88d   &                         &                        &10.2&26.6&0.02   &  29.83 & 0.40  & 22.54 \\
90                 &        3c           & HCG90a   &                    49.1 &                  39.9  &9.8&1.8&0.68   &  21.88   & 0.53  & 23.38 \\
\multirow{3}{*}{91}&  \multirow{3}{*}{2} & HCG91a   &   \multirow{3}{*}{51.3} & \multirow{3}{*}{47.0}  &30.7&1.9&0.53   &  58.82  & 1.03  & 51.61 \\
                   &                     & HCG91b   &                         &                        &31.0&1.6&-0.06  &  53.58  & 1.76  & 34.48 \\
                   &                     & HCG91c   &                         &                        &31.4&26.5&0.01   &  39.21 & 0.62  & 24.11 \\
93                 &        3c           & HCG93b   &                    59.7 &                  53.3  &21.7&0.4&0.32   &  42.76  & 0.29  & 30.33 \\
96                 &        2            & HCG96a   &                    26.9 &                  18.3  &18.3&5.4&0.28  &  75.73   & 0.64  & 37.86 \\
\bottomrule
\end{tabular}
\tablefoot{Columns: (1) HCG ID number, (2) Phase of the group, (3) member galaxies considered in this work, (4--5) Major and minor axes of the beam, (6) linear resolution of the observations, (7) \HI\ column density sensitivity limit over 20 $\mathrm{km~s^{-1}}$, (8) \HI\ deficiency from \citet{2023A&A...670A..21J}, (9) \HI\ diameter, (10) Error in \HI\ diameter, (11) optical diameter from the HyperLeda database.}
\end{table*}

\section{Methodology}\label{methodology}
We estimate the size of the \HI\ disk at 1 $\mathrm{M_{\odot}~pc^{-2}}$ using the following procedures. First, we convert the moment-zero maps to column density maps using:
\begin{equation}
N_{\mathrm{HI}} = 1.104 \times 10^{21} \times \frac{I_{\mathrm{HI}}}{\left( \mathrm{B}_{\mathrm{maj}} \times \mathrm{B}_{\mathrm{min}} \right)},
\end{equation}
where $I_{\mathrm{HI}}$ is the integrated intensity in $\mathrm{Jy~km~s^{-1}}$, $B_{\mathrm{maj}}$ and $B_{\mathrm{min}}$ are the major and minor axes of the beam in arcseconds. After that, we convert the column density maps to surface density maps using the following formula:
\begin{equation}
    \Sigma_{\mathrm{HI}}~[\mathrm{M_{\odot}~pc^{-2}}] = N_{\mathrm{HI}}~[cm^{-2}]~/~1.248~\times~10^{20}.
\end{equation}
We use Matplotlib to extract collections of points that make up the contour line corresponding to 1 $\mathrm{M_{\odot}~pc^{-2}}$. These are sets of (x,y) coordinates obtained by extracting the vertices of the contour. Then, we fit these points using the ``Numerically stable direct least squares fitting of ellipses'' described in \citet{oy1998NumericallySD}. This method is robust to noise and outliers and works well even when the contour covers only part of the ellipse. The fitting is noniterative, which makes it fast. We prefer to use this direct ellipse fitting of the contour at 1 $\mathrm{M_{\odot}~pc^{-2}}$ as opposed to, e.g., fitting Gaussians to the \HI\ surface density distribution \citep{2022MNRAS.512.2697R} as the latter requires an assumption on the gas distribution. The fitting technique is already described in detail by \citet{oy1998NumericallySD}. However, because this is the first application of the method to measure $\mathrm{D_{HI}}$, we provide a description in Appendix~\ref{app:fitting-procedure} for completeness and to ensure that this paper is self-contained. To account for the effect of the beam, we subtract the effective beam area in quadrature as done in \citet{2016MNRAS.460.2143W} using the following equation:

\begin{equation}
D_{\mathrm{HI}} = \sqrt{D_{\mathrm{HI,obs}}^{2} - (B_{\mathrm{min}} \times B_{\mathrm{maj}}}),
\end{equation}
where $D_{\mathrm{HI,obs}}$ is the measured diameter from the ellipse fitting described in Appendix~\ref{app:fitting-procedure}; \(B_{\mathrm{min}}\) and \(B_{\mathrm{maj}}\) are the minor and major axes of the beam, respectively. For three galaxies in the HCGs sample, the measured diameters are smaller than the beam size. For these, we use their major-axis beam size ($B_{\rm maj}$) as upper limits to their \HI\ diameters. In addition, \HCGUpperLimitCount\ HCG members in the \citet{2023A&A...670A..21J} sample are not detected in \HI. These galaxies are listed in Appendix~\ref{app:upperlimits} (Table~\ref{table:upperlimits}). We retain these as left-censored data, assigning each an upper limit on $D_{\rm HI}$ equal to their $B_{\rm maj}$ and analyze their distribution using survival-analysis techniques that account for censored values. Following the approach adopted by \citet{2024A&A...687A.244H}, we use Kaplan--Meier estimators \citep{Kaplan01061958} to determine median values and apply the Gehan generalized Wilcoxon test \citep{1965BiometrikaGehan} to assess the differences between the HCG sample and the comparison samples. 
The \HI\ mass is obtained from the moment-zero maps using the Source Finding Application \citep[SoFiA,][]{2015MNRAS.448.1922S, 2021MNRAS.506.3962W}, including emission beyond $\mathrm{\Sigma_{HI} = 1~M_{\odot}~pc^{-2}}$, using
\begin{equation}
M_{\mathrm{HI}}[M_{\odot}] = 2.36 \times 10^{5} \, D^{2} \times \sum_{i} S_{i}\,\Delta v,
\end{equation}

where $\sum_{i}(S_{i}\Delta v)$ is the integrated intensity in Jy\,km\,s$^{-1}$ and $D$ is the distance to the galaxy in Mpc. For the HCGs we adopt Cosmicflows-3 flow-model distances following \citet{2023A&A...670A..21J}. For the AMIGA galaxies we adopt the distances of \citet{2018A&A...609A..17J}, derived from the multi-attractor flow model of \citet{2000ApJ...529..786M}. Recomputing the AMIGA distances with the Cosmicflows-3 model changes the distances by only $\sim$1\% in the median and the measured truncation by less than $0.002$ dex, which is negligible compared to the observed scatter. 
\subsection{Error analysis}\label{sub:error-analysis}
We estimate the uncertainty on the \HI\ mass using the formula below. 
\begin{equation}
    \sigma_{\mathrm{M_{HI}}} = 2.356 \times 10^{5}~D^{2}~\sigma_{ch}~\Delta v~\sqrt{N_{eff}}, 
\end{equation}
where $D$ is the distance in Mpc, $\sigma_{ch}$ is the rms noise level of the data cubes in $\mathrm{Jy~beam^{-1}}$ estimated using SoFiA, $\Delta v$ is the channel width in $\mathrm{km~s^{-1}}$, and $N_{\mathrm{eff}}$ is the effective number of independent beam-channel elements contributing to the integrated flux defined as: 

\begin{equation}\label{eq:neff}
    N_{\mathrm{eff}} = \frac{A_{\mathrm{pix}}}{A_{\mathrm{beam}}}
    \sum_{(x,y)\in M} N_{\mathrm{ch}}(x,y)\,,
\end{equation}

where $A_{\mathrm{beam}}$ and $A_{\mathrm{pix}}$ are the areas of the beam and a pixel, respectively; $N_{\mathrm{ch}}(x,y)$ is the number of spectral channels included within the SoFiA mask at pixel position $(x,y)$; $\sum\limits_{\substack{(x,y)\in M}}$ denotes the summation over all pixels $(x,y)$ within the SoFiA mask $M$.  
Note that the error does not take into account the uncertainty in the distance estimate. \\ \indent 
We estimate the uncertainty on $D_{\rm HI}$ using a Monte Carlo approach that 
accounts for the spatially correlated noise structure imposed by the synthesized 
beam. The surface-density uncertainty at each pixel is derived by propagating 
the per-channel noise through the moment-zero integration. The flux uncertainty 
at pixel $(x,y)$ is $\sigma_{\rm ch}\,\Delta v\,\sqrt{N_{\rm ch}(x,y)}$, which 
we convert to surface-density units via
\begin{equation}\label{eq:sigma_surf}
    \sigma_{\Sigma}(x,y) = \frac{C_{\rm conv}\,\sigma_{\rm ch}\,\Delta v\,
    \sqrt{N_{\rm ch}(x,y)}}{\Omega_{\rm beam}}\,,
\end{equation}
where $C_{\rm conv} = 1.104 \times 10^{21} / 1.248 \times 10^{20}$ is the 
combined unit-conversion factor from $\mathrm{Jy\,beam^{-1}\,km\,s^{-1}}$ to 
$M_{\odot}\,\mathrm{pc}^{-2}$, and $\Omega_{\rm beam} = (\pi/4\ln 2)\,
B_{\rm maj}\,B_{\rm min}$ is the beam solid angle in $\mathrm{arcsec}^{2}$.

To simulate the effect of measurement noise on the derived diameters while 
preserving the spatial correlation structure of the data, we generate 
beam-correlated noise realizations as follows:
\begin{enumerate}
    \item Generate a white-noise field $w(x,y) \sim \mathcal{N}(0,1)$ with 
          the same dimensions as the surface-density map.
    \item Convolve $w(x,y)$ with a Gaussian kernel matching the synthesized 
          beam to produce a spatially correlated noise field $w_{\rm corr}(x,y)$.
    \item Scale the correlated noise field to match the local uncertainty:
          \begin{equation}
              \eta(x,y) = \sigma_{\Sigma}(x,y) \times 
              \frac{w_{\rm corr}(x,y)}{\sigma(w_{\rm corr})}\,,
          \end{equation}
          where $\sigma(w_{\rm corr})$ is the standard deviation of the 
          convolved white-noise field.
\end{enumerate}

Because the observed surface-density map $\Sigma_{\rm HI}(x,y)$ already contains 
one realization of measurement noise, adding a second noise field would inflate 
the total noise by a factor of $\sqrt{2}$. To take this into account, we first construct a 
smoothed estimate of the underlying signal by applying a median filter (with a 
kernel size of three beam widths) to the original map, denoted 
$\tilde{\Sigma}_{\rm HI}(x,y)$. Each Monte Carlo realization is then formed as
\begin{equation}
    \Sigma_{i}(x,y) = \tilde{\Sigma}_{\rm HI}(x,y) + \eta_{i}(x,y)\,,
\end{equation}
where $\eta_{i}$ is the $i$-th beam-correlated noise realization. This ensures 
that each perturbed map has a noise level consistent with the observations.

We generate 500 such realizations. For each, we extract the 
$1\,M_{\odot}\,\mathrm{pc}^{-2}$ isophotal contour, apply bootstrap resampling 
of its vertices, and fit an ellipse following the same procedure used for the 
unperturbed map. The reported uncertainty $\sigma_{D_{\rm HI}}$ is the standard 
deviation of the diameters measured across all 500 realizations.
\subsection{Do AMIGA and HCGs follow the \HI\ size--mass relation?}
The \HI\ size--mass relation is known to be remarkably tight across samples spanning a wide range of galaxy types and environments \citep[e.g.,][]{B1997, 2016MNRAS.460.2143W, 2019MNRAS.490...96S, 2022MNRAS.512.2697R}. This has been interpreted as reflecting broadly similar \HI\ surface-density profiles in the outer disks, so that changes in \HI\ extent are closely coupled to changes in total \HI\ mass \citep{2016MNRAS.460.2143W}. We therefore test whether the extreme environments of AMIGA and HCGs (stringent isolation versus frequent strong interactions) produce measurable systematic deviations from this established relation. To do this, we fit the size--mass relation, log($D_{\mathrm{HI,obs}}$) vs log($M_{\rm HI}$), for the resolved AMIGA and HCGs using a Bayesian linear regression model that accounts for measurement uncertainties, and test whether the resulting slopes and intercepts are statistically indistinguishable from those obtained from a fit to the literature sample, consisting of the compiled data from \citet{2016MNRAS.460.2143W} and the MeerKAT International GHz Tiered Extragalactic Exploration Survey \citep[MIGHTEE,][]{2016mks..confE...6J} data from \citet{2022MNRAS.512.2697R}. If AMIGA and HCGs are consistent with the established size--mass relation, we can use it to infer $D_{\mathrm{HI,obs}}$ for the AMIGA galaxies that have only single-dish \HI\ measurements, which will increase our isolated-galaxy sample.
\\ \indent We model the mass-size relation as:
\[
\log D_{\mathrm{HI}} = m \log M_{\mathrm{HI}} + b + \epsilon,
\]
where $\epsilon$ represents the intrinsic scatter in $\log D_{\mathrm{HI}}$ and is assumed to follow a normal distribution with zero mean and standard deviation $\sigma_{\mathrm{int}}$. The free parameters of the model are the slope $m$, intercept $b$, and the logarithm of the intrinsic scatter, $\ln \sigma_{\mathrm{int}}$.

We assume Gaussian measurement errors in $\log D_{\mathrm{HI}}$ and adopt a standard likelihood that combines the measurement error and intrinsic scatter in quadrature:
\[
\mathcal{L}_{\mathrm{det}} = \prod_i \frac{1}{\sqrt{2\pi (\sigma_{\mathrm{int}}^2 + \delta_i^2)}} \exp\left[ -\frac{(y_i - m x_i - b)^2}{2(\sigma_{\mathrm{int}}^2 + \delta_i^2)} \right],
\]
where $\delta_i$ is the measurement uncertainty in $\log D_{\mathrm{HI}}$.

\begin{figure}
    \centering
    \begin{tabular}{c}
       \includegraphics[scale=0.45]{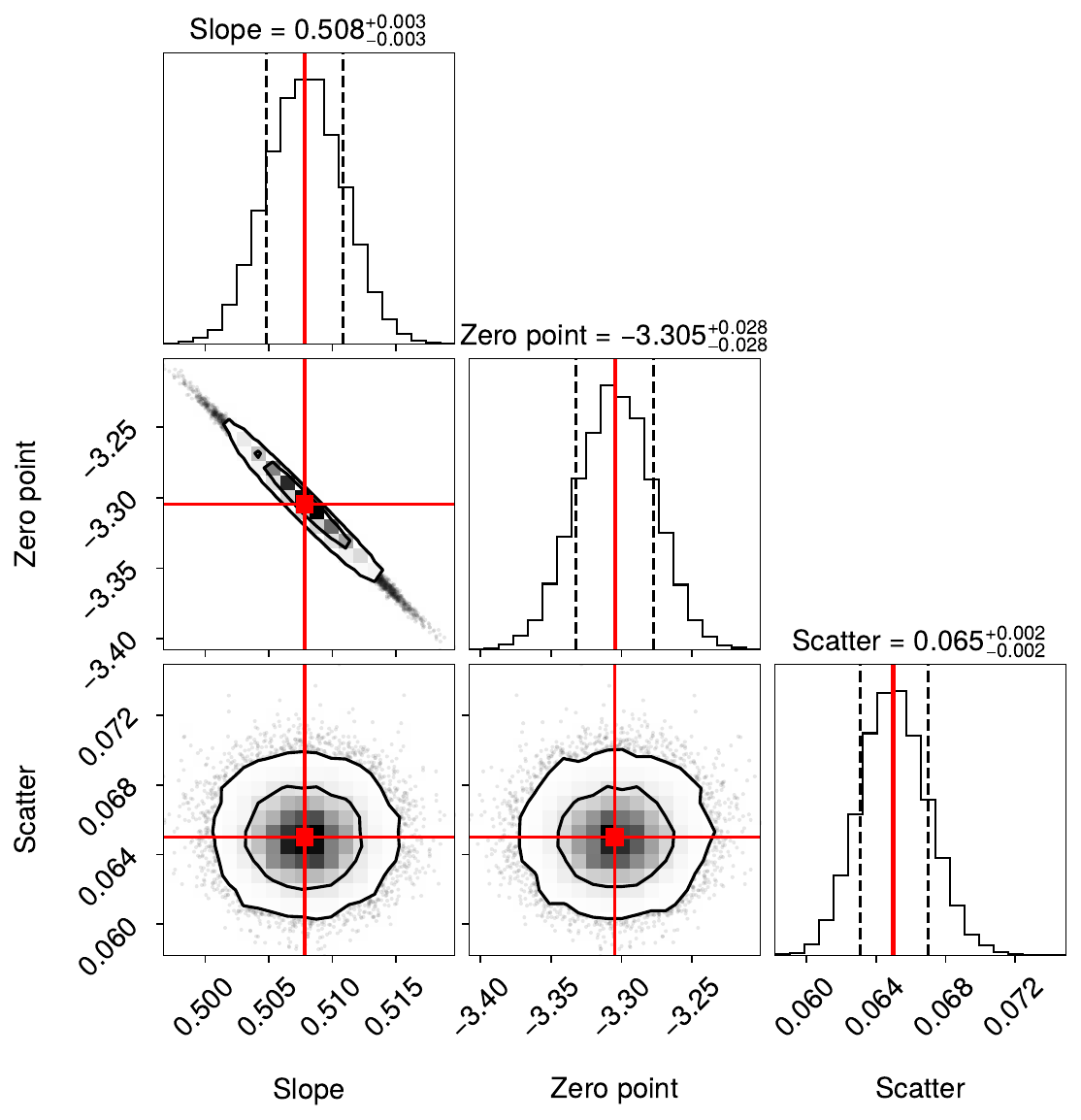}
       \end{tabular}
	    \caption{Posterior distributions of the slope, intercept, and intrinsic scatter of the \HI\ mass--size relation fitted to the combined sample of 727 galaxies (resolved AMIGA and HCGs from this work, together with the literature data of \citet{2016MNRAS.460.2143W} and \citet{2022MNRAS.512.2697R}). This is the calibration we use to infer $D_{\mathrm{HI}}$ for the single-dish AMIGA galaxies that lack resolved \HI\ maps. The red lines indicate the posterior median values. The contours enclose the 68\% and 95\% credible regions.}
       \label{fig:corner-plot}
\end{figure}

We sample the posterior distribution of the parameters using the Python \texttt{emcee} ensemble Markov Chain Monte Carlo (MCMC) sampler. To guide the initialization of the chains, we first compute an ordinary least squares fit to the data points. We then initialize the MCMC walkers near this solution, and let the chains evolve for 4000 steps. We discard the first 1000 steps as burn-in to ensure convergence, and thin the chain by a factor of 15 to reduce autocorrelation between samples. Figure~\ref{fig:corner-plot} shows the resulting posterior distributions of the slope, intercept, and intrinsic scatter of the combined sample (resolved AMIGA, HCGs, MIGHTEE, and the Wang+16 compilation), which we adopt to infer $D_{\mathrm{HI}}$ for the unresolved single-dish AMIGA galaxies, obtained from the MCMC sampling.

Table~\ref{table:consistency_fits} shows the fitting results for the individual samples and the combined data set. 
\begin{table}
\centering
\caption{Bayesian fit parameters for the \HI\ mass--size relation.\label{table:consistency_fits}}
\begingroup
\renewcommand{\arraystretch}{1.2}
\resizebox{\columnwidth}{!}{%
\begin{tabular}{lcccc}
\toprule \toprule
Sample & $m$ & $b$ & $\sigint$ Scatter & $N$ \\
\midrule
AMIGA & $0.455^{+0.035}_{-0.035}$ & $-2.732^{+0.346}_{-0.347}$ & $0.080^{+0.012}_{-0.010}$ & 35 \\
\addlinespace[0.35em]
HCGs & $0.509^{+0.025}_{-0.027}$ & $-3.265^{+0.250}_{-0.239}$ & $0.077^{+0.009}_{-0.008}$ & 53 \\
\addlinespace[0.35em]
AMIGA + HCGs & $0.496^{+0.020}_{-0.019}$ & $-3.144^{+0.185}_{-0.187}$ & $0.078^{+0.007}_{-0.006}$ & 88 \\
\addlinespace[0.35em]
MIGHTEE & $0.503^{+0.008}_{-0.008}$ & $-3.268^{+0.074}_{-0.073}$ & $0.053^{+0.003}_{-0.003}$ & 204 \\
\addlinespace[0.35em]
Wang2016 & $0.506^{+0.004}_{-0.004}$ & $-3.291^{+0.031}_{-0.031}$ & $0.063^{+0.002}_{-0.002}$ & 435 \\
\addlinespace[0.35em]
MIGHTEE + Wang2016 & $0.504^{+0.003}_{-0.003}$ & $-3.276^{+0.025}_{-0.026}$ & $0.060^{+0.002}_{-0.002}$ & 639 \\
\addlinespace[0.35em]
Combined sample & $0.508^{+0.003}_{-0.003}$ & $-3.305^{+0.028}_{-0.027}$ & $0.065^{+0.002}_{-0.002}$ & 727 \\
\bottomrule
\end{tabular}%
}
\endgroup
\tablefoot{Columns: slope~$m$, intercept~$b$, intrinsic scatter~$\sigint$, and sample size~$N$.
Quoted uncertainties are the central 68\% credible intervals (16th--84th percentiles).}
\end{table}
\begin{figure*}
    \centering
    \begin{tabular}{c}
	    \includegraphics[scale=0.65]{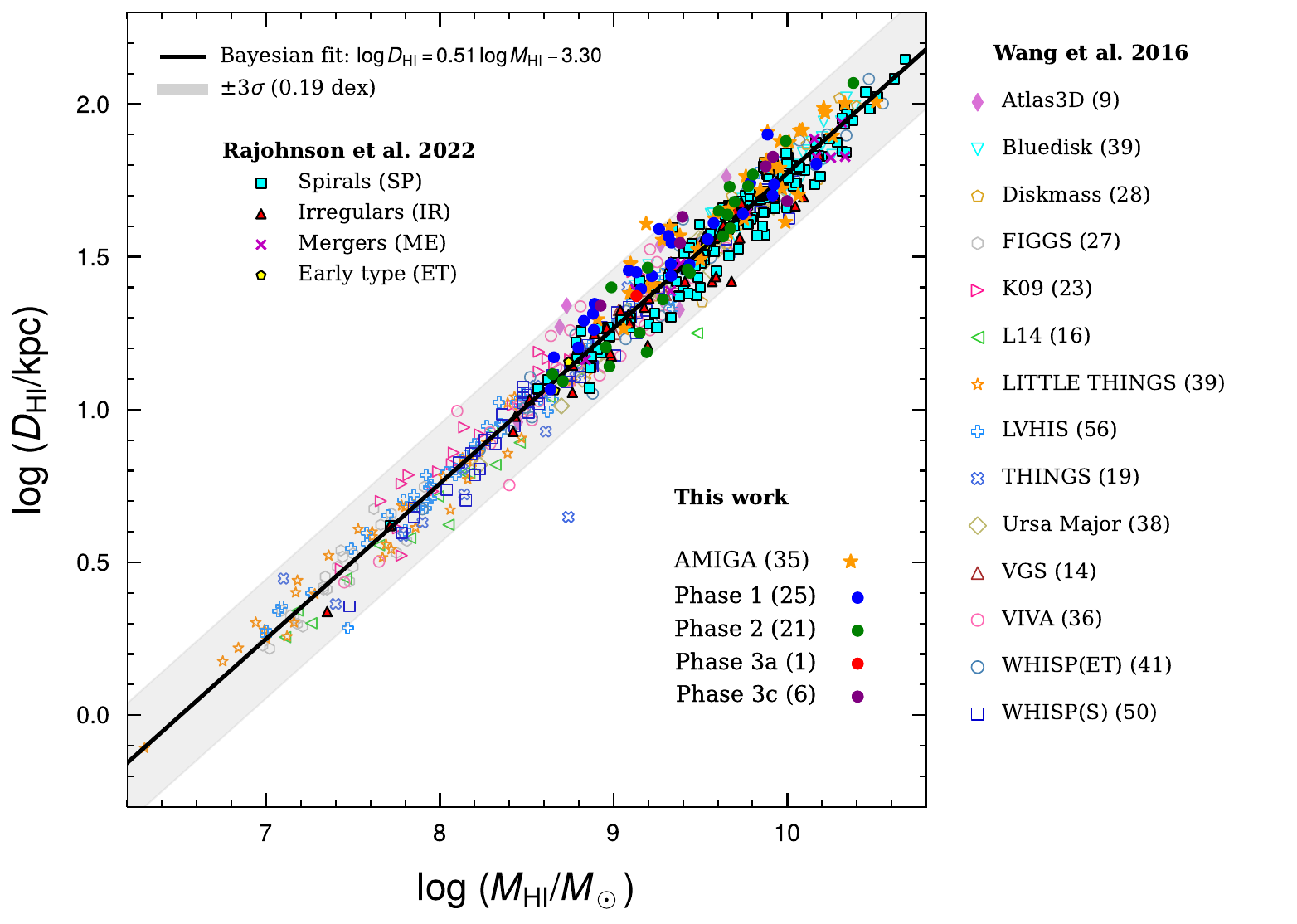}
       \end{tabular}
	\caption{\HI\ mass vs \HI\ diameter. The filled stars and circle symbols are the results from this work and indicate galaxies from AMIGA and HCGs, respectively. The rest of the symbols indicate results from \citet{2016MNRAS.460.2143W} and \citet{2022MNRAS.512.2697R}, which use data from the MIGHTEE survey. The solid black line shows the best-fit relation for the combined sample of all galaxies from this work and the literature data. The gray shaded area indicates the $\pm3\sigma$ scatter around the fit.}
       \label{fig:dhi-mhi}
\end{figure*}
We take the median values of the posterior distributions for the slopes, intercepts, and intrinsic scatters as the best-fitting values. We define the uncertainties around each parameter as the 16th and 84th percentiles of the posterior distribution. As shown in Table~\ref{table:consistency_fits}, the AMIGA values slightly deviate from the rest of the samples. In fact, the AMIGA (resolved) slope is about $\AmigaSlopeTension\sigma$ below that of the MIGHTEE+Wang2016 sample; however, because its uncertainty is an order of magnitude larger than the literature uncertainty, there is still a strong overlap between the two posteriors. The AMIGA fit yields an intercept that is about $\AmigaInterceptTension\sigma$ above the literature intercept. To test whether the difference between the best-fitting values is statistically significant or not, we analyze the posterior distribution of the parameter differences $\Delta m$ (difference in slope between the sample being compared) and $\Delta b$ (difference in intercept). If the samples being compared are drawn from the same underlying relation, the posterior distributions of $\Delta m$ and $\Delta b$ should be statistically consistent with zero. We quantify the agreement by expressing the median parameter difference in units of its posterior uncertainty. For AMIGA, the slope and intercept differ from the literature values by only $\AmigaSlopeTension\sigma$ and $\AmigaInterceptTension\sigma$. This indicates that the AMIGA size--mass relation is mildly offset from that of the literature but not statistically significant. For HCGs, the differences are within $\HcgsSlopeTension\sigma$ and $\HcgsInterceptTension\sigma$ for the slope and intercept, respectively, indicating a strong agreement with the literature. \\ \indent 
The slope and intercept differences for AMIGA should not be interpreted as two independent indications of a real offset from the literature relation, since these two parameters are strongly anticorrelated in the fit. A slightly shallower slope leads to a higher intercept, especially for a sample that has only \AmigaN\ galaxies and spans a relatively limited range in \HI\ mass. Therefore, the mild deviation seen for AMIGA is most likely driven by small-number statistics and by the slope--intercept degeneracy, rather than by a genuine physical difference in the underlying \HI\ mass--size relation. This interpretation is supported by the combined AMIGA+HCGs comparison, for which the differences with respect to the literature values become much smaller. Any environmental imprint would most likely be noticeable in the HCGs vs literature comparison, yet the HCGs posterior is statistically indistinguishable from the MIGHTEE+Wang2016 reference. A plausible explanation follows from the slope of the relation itself. A slope of $m\simeq0.5$ ($D_{\mathrm{HI}}\propto M_{\mathrm{HI}}^{0.5}$, equivalently $M_{\mathrm{HI}}\propto\langle\Sigma_{\mathrm{HI}}\rangle\,D_{\mathrm{HI}}^{2}$) implies that all galaxies share a roughly universal mean \HI\ surface density within $D_{\mathrm{HI}}$, independent of mass or environment. Tidal stripping removes the low-column-density gas from the disk outskirts, lowering $D_{\mathrm{HI}}$ and $M_{\mathrm{HI}}$ together (with $M_{\mathrm{HI}}\propto D_{\mathrm{HI}}^{2}$) while leaving this characteristic mean surface density essentially unchanged. Galaxies, therefore, move along the relation rather than perpendicular to it. This interpretation is consistent with the finding of \citet{2016MNRAS.460.2143W}, who showed that even galaxies affected by ram pressure in the Virgo cluster (e.g., NGC\,4402) deviate by as little as 0.01\,dex from the mean relation. Only galaxies with kinematical abnormalities (counterrotating disks,
ongoing mergers) or extreme H\,{\sc i} deficiency were found to be $>3\sigma$ outliers \citep{2016MNRAS.460.2143W}.\\ \indent  The MIGHTEE sample (N\,=\,\MighteeN) exhibits the lowest intrinsic scatter, $\sigma_{\mathrm{int}}=\MighteeScatterDex$\,dex. This is the first blind, homogeneous H\,{\sc i} survey used to study the size--mass relation. The data were obtained with a single instrument (MeerKAT) and processed through a uniform pipeline. The lower scatter compared to the AMIGA/HCG samples and the \citet{2016MNRAS.460.2143W} compilation can be attributed in large part to the absence of systematic uncertainties introduced by combining heterogeneous data sets from different telescopes, with different angular resolutions, and sensitivity limits. \\ \indent  
We conclude that the size--mass relations for AMIGA and HCGs are statistically consistent with the MIGHTEE+Wang2016 size--mass relation, despite the mild offset seen for AMIGA when it is fitted alone as demonstrated previously. We therefore adopt the fit to the combined sample (MIGHTEE + Wang2016 + AMIGA + HCGs) as our \HI\ size--mass relation. \\ \indent 
We overplot the best-fit result on the size--mass relation in Fig.~\ref{fig:dhi-mhi} for the combined sample. We use this relation to infer the $D_{\mathrm{HI}}$ of the AMIGA galaxies that have only single-dish \HI\ measurements, thereby extending our sample of isolated galaxies. For this, we use the single-dish sample of \citet{2018A&A...609A..17J}, selecting the 372 CIG galaxies with measured \HI\ masses and optical diameters and excluding the five sources flagged by the authors as exceptionally low in \HI\ mass, which appear as outliers in the size-mass and \HI\ mass vs. $B$-band luminosity relations.   
\section{Results}\label{results}
\begin{figure*}
    \centering
    \begin{tabular}{ccc}
       \includegraphics[scale=0.47]{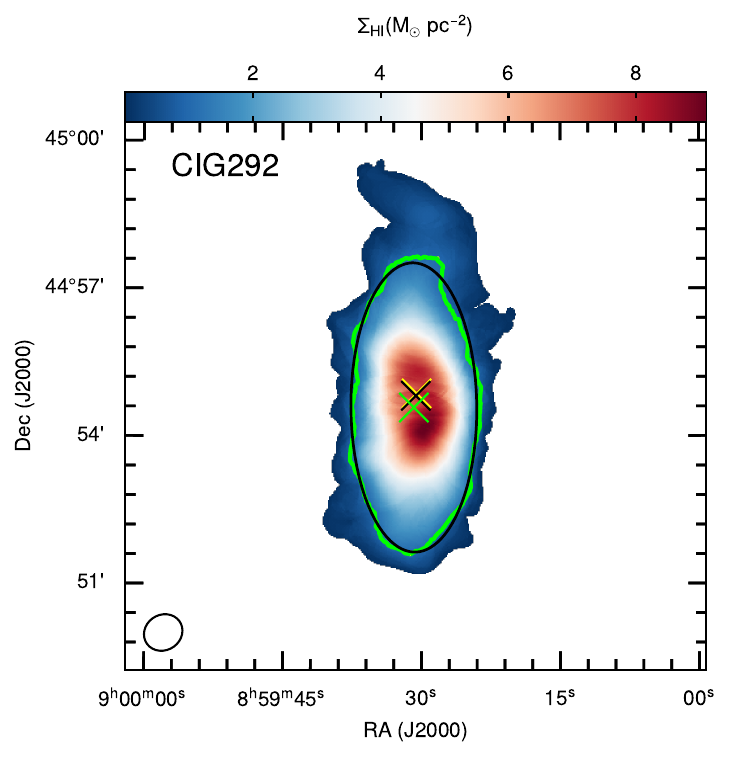} &
       \includegraphics[scale=0.47]{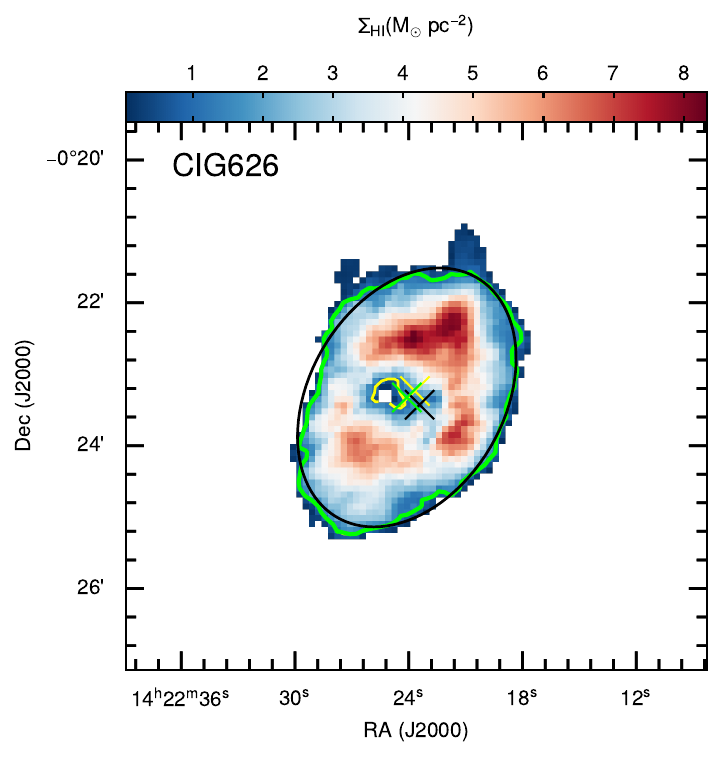}&
       \includegraphics[scale=0.47]{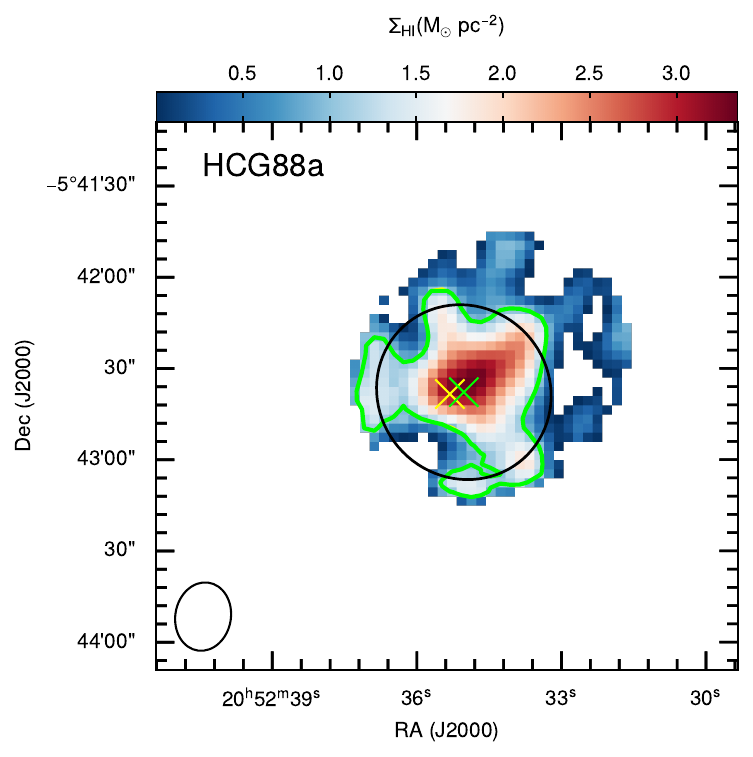} 
       \end{tabular}
	\caption{Ellipse fits to the \HI\ iso-surface density contour at \(1\,\mathrm{M_\odot\,pc^{-2}}\) for three galaxies. The yellow line (mostly hidden behind the green line) represents the \HI\ surface density contour at \(1\,\mathrm{M_\odot\,pc^{-2}}\). The green line shows the contour vertices that were used for the ellipse fitting, which excludes outliers. The black line represents the ellipse fitting to the vertices. The yellow crosses indicate the optical center of the galaxies, the black crosses indicate the kinematical center, and the green crosses represent the center of the fitted ellipses. The rest of the figures are shown as online material.}
       \label{fig:ellipse-fit}
\end{figure*}

In Fig.~\ref{fig:ellipse-fit}, we show examples of the ellipse fits to the \HI\ iso-surface density contour at \(1\,\mathrm{M_\odot\,pc^{-2}}\) for three galaxies. The ellipse fits for the full sample are available as online material.
\subsection{\HI\ disk truncation across HCG phases relative to isolated galaxies}\label{sub:diameter-optical}
\begin{figure*}
    \centering
    \begin{tabular}{c c}
       \includegraphics[scale=0.4]{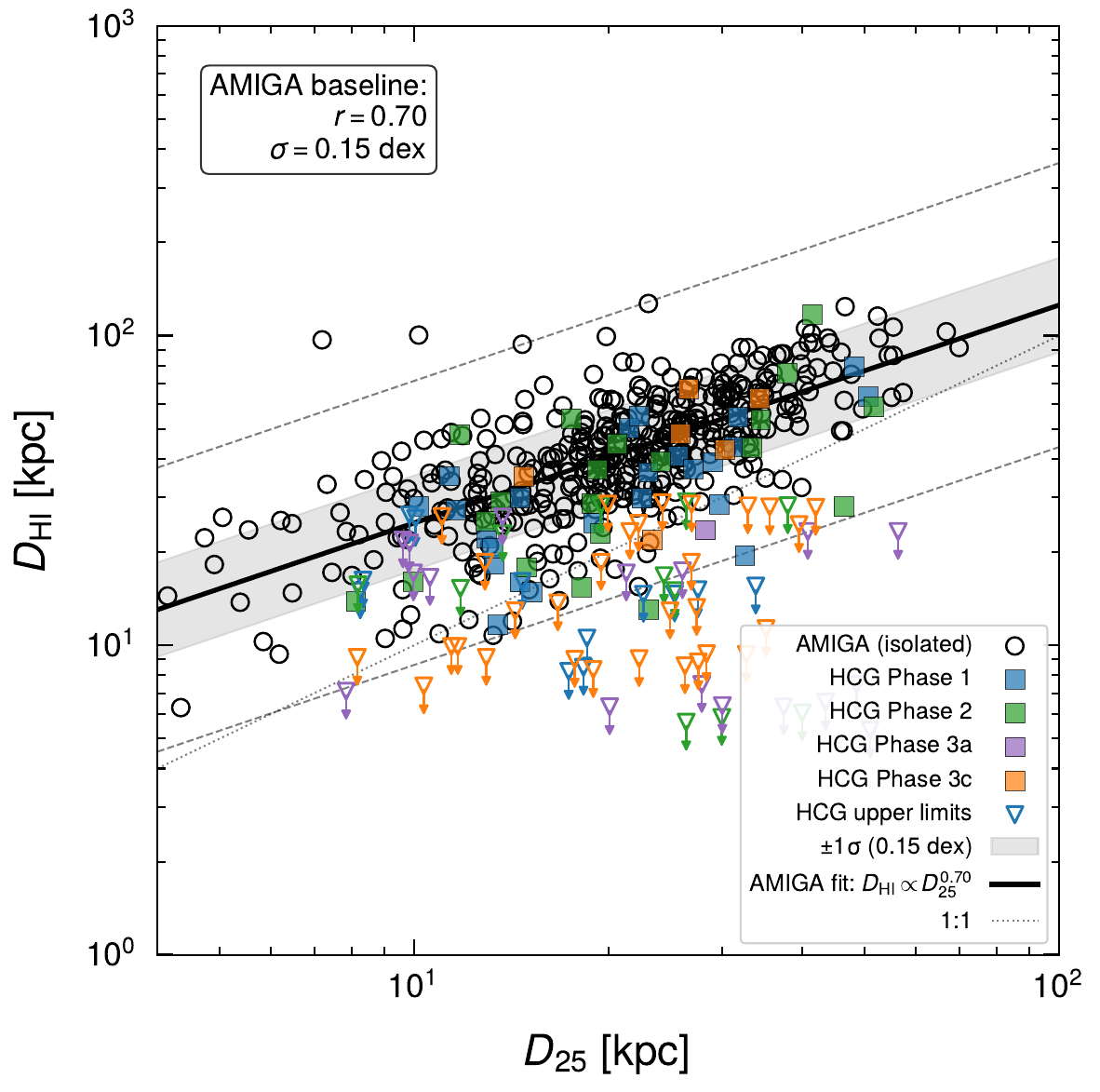} & 
       \includegraphics[scale=0.4]{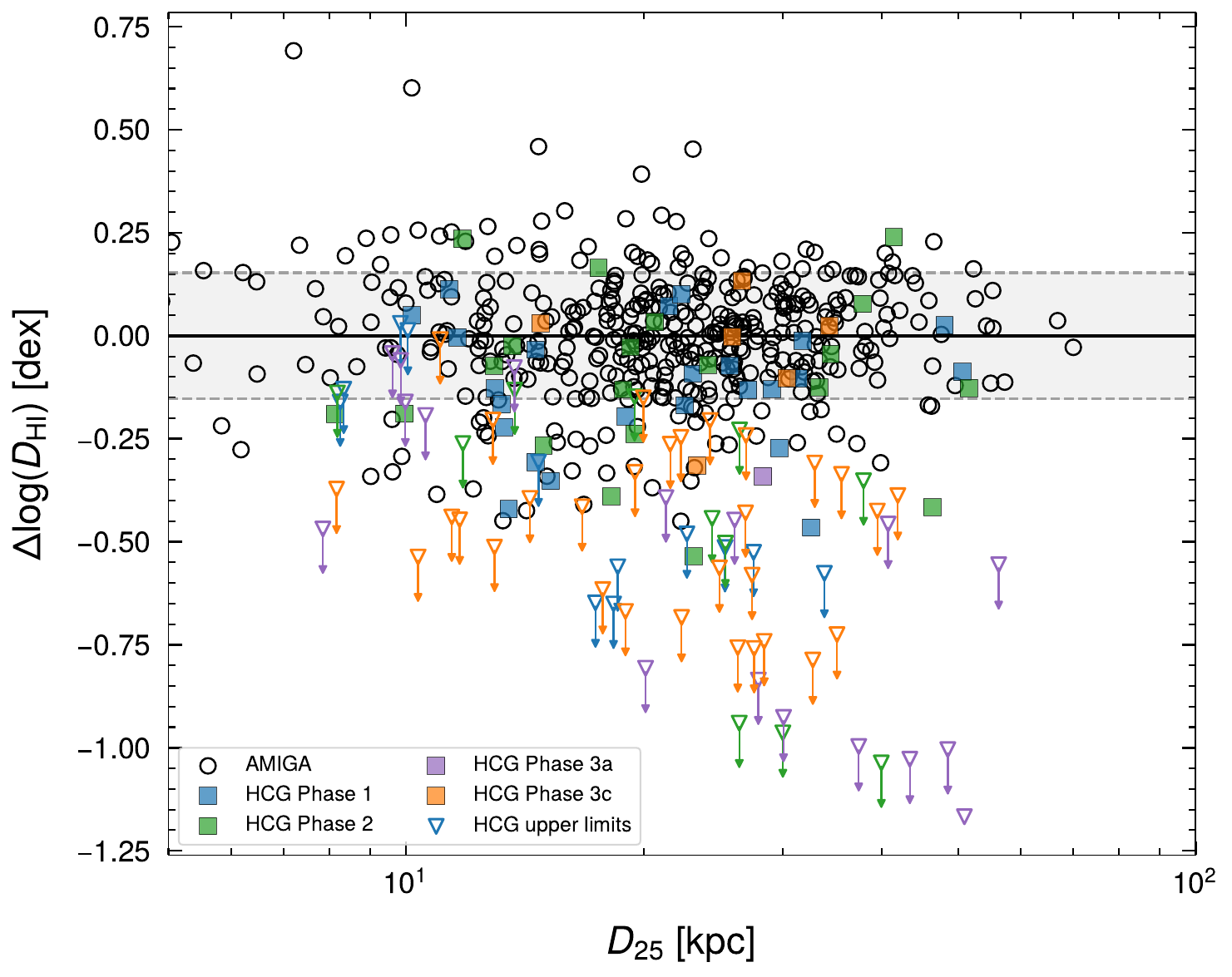}
       \end{tabular}
       \caption{$D_{\rm HI}$ versus $D_{25}$ in logarithmic space (left panel) and $\Delta \log (D_{\rm HI})$ versus $D_{25}$ (right panel). Open circle symbols indicate the combined AMIGA (isolated) galaxies and square symbols represent HCG galaxies. The solid line in the left panel is a Bayesian fit to the AMIGA sample (Eq.~\ref{eq:baseline}). The solid line in the right panel represents $\Delta \log (D_{\rm HI})$ = 0 (the AMIGA reference). The gray area marks the $\pm 1\sigma$ scatter band. The dashed lines indicate the $\pm 3\sigma$ bounds. The dotted line is a 1:1 relation. }
       \label{fig:diameter_correlation}
\end{figure*}

\begin{figure*}
    \sidecaption
       \includegraphics[width=12cm]{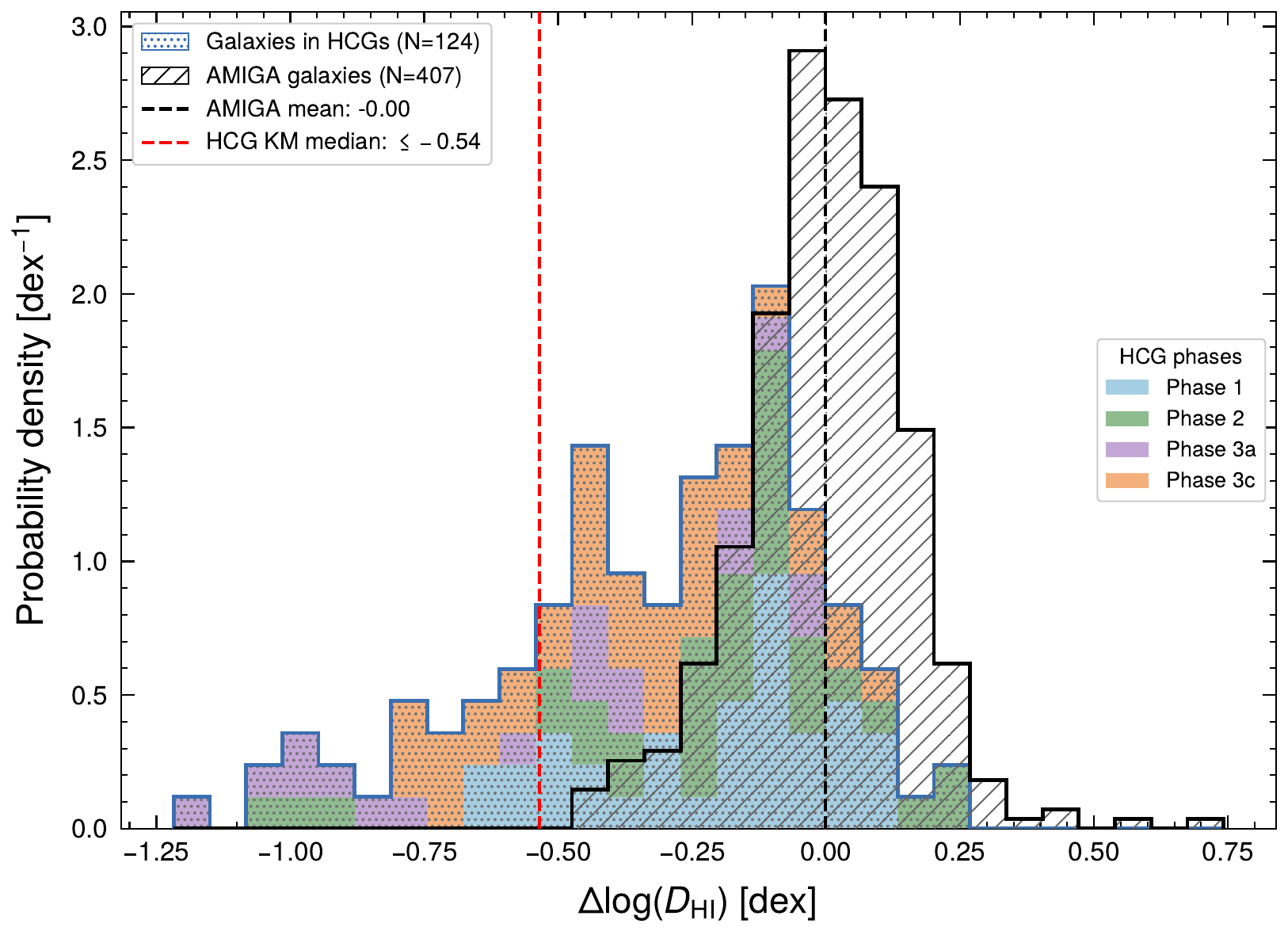}
       \caption{Probability-density distributions of the residuals $\Delta\log(D_{\rm HI})$ for AMIGA (hatched, $N=407$) and HCG (filled, color-coded by evolutionary phase) galaxies. The HCG sample includes the $70$ \HI\ nondetected members as beam-size upper limits (binned at their limit value; Appendix~\ref{app:upperlimits}), giving $N=124$. The black dashed line marks the AMIGA mean and the red dashed line the Kaplan--Meier median of the HCG residuals (upper limits included). More than half of the HCG members are upper limits, so the Kaplan--Meier median lies in the censored tail, below the peak formed by the detections; the marked value is its rigorous upper bound. The histogram bars are color-coded by evolutionary phase within each bin. }
       \label{fig:residuals_hist}
\end{figure*}

In this study, we adopt $D_{\mathrm{25}}$ as the apparent diameter of a galaxy measured at the isophotal level of 25 $B$-mag arcsec$^{-2}$ taken from the HyperLeda database \citep{2003A&A...412...45P}. While previous studies used the \HI\ to optical diameter ratio as a tracer of the gas removal process in the outer disk of galaxies \citep{1988A&AS...72..427W,1994AJ....107.1003C,B1997,2002A&A...390..829S,2007MNRAS.378..137S,2009AJ....138.1741C,2016MNRAS.460.2143W, 2017ASSL..434..209B, 2022MNRAS.510.1716R}, in this paper we adopt the correlation and residuals methodology of \citet{1997ApJ...490..166P}. First, we fit the $\log(D_{25})$--$\log(D_{\rm HI})$ relation (see Eq.~\ref{eq:sizesize}) of the combined AMIGA sample, i.e., the 35 resolved galaxies plus the 372 galaxies with single-dish measurements (after excluding those that have no $D_{\mathrm{25}}$ measurement). 
\begin{equation}\label{eq:sizesize}
    \log D_{\rm HI} \;=\; \alpha\,\log D_{25} + \beta,
\end{equation} 
This fit defines the expected $D_{\rm HI}$ for an isolated galaxy of a given optical diameter, and serves as our reference baseline. Our aim is to quantify how the HCGs and different samples in the literature deviate from this adopted baseline. Therefore, for each galaxy in both the isolated and interacting samples, we compute the residual \begin{equation} \Delta \log(D_{\rm HI}) = \log(D_{\rm HI}^{\rm obs}) - \log(D_{\rm HI}^{\rm exp}),\end{equation} where $D_{\rm HI}^{\rm exp}$ is the \HI\ diameter predicted by the AMIGA baseline at that galaxy's $D_{25}$. A negative residual means the galaxy's \HI\ disk is smaller than expected for an isolated galaxy of the same optical size (i.e., truncated), while a positive residual indicates an \HI\ disk that is more extended than expected. We then compare the residual distributions of the samples being compared using non-parametric tests (Mann--Whitney $U$, Kolmogorov--Smirnov) and quantify the mean offset between them. We break down the HCG residuals by evolutionary phase to examine any trend between the degree of \HI\ truncation and evolutionary sequence. We prefer this residual-based approach over the traditional $D_{\rm HI}/D_{25}$ ratio for the following reason \citep[see also][]{1997ApJ...490..166P}. The relationship between $D_{\rm HI}$ and $D_{25}$ is nonlinear as the \HI\ diameter scales as $D_{\rm HI} \propto D_{25}^{\alpha}$, where $\alpha \neq 1$ in general. So, when $\alpha \neq 1$, the ratio $D_{\rm HI}/D_{25}$ carries a systematic dependence on $D_{25}$ itself, so that larger galaxies will tend to have systematically different ratios than smaller ones purely because of the nonlinear scaling, not because of any environmental effect. Comparing ratios between two samples that span different ranges of $D_{25}$ can therefore introduce a spurious trend (or mask a real one). Our approach removes this size-dependent bias. It ensures that any offset we detect between the samples' residual distributions reflects actual \HI\ truncation (or enhancement) at fixed optical size, rather than an artifact of comparing a nonlinear ratio across samples with different diameter distributions. Figure~\ref{fig:diameter_correlation} presents the $D_{\rm HI}$--$D_{25}$ plane for the AMIGA and HCG samples, together with the adopted AMIGA baseline derived below and the corresponding residuals as a function of $D_{25}$.
\\ \indent The choice of the linear estimator to use to fit the $\log(D_{25})$--$\log(D_{\rm HI})$ relation is important. This is because if the residuals of the adopted fit correlate with $\log D_{25}$, using that fit as a baseline against which the $\Delta \log(D_{\rm HI})$ of other samples are measured will inherit a size-dependent bias that is an artifact of the estimator rather than a property of the data. Thus, we tested several fitting methods to use for Eq.~\ref{eq:sizesize}, including ordinary least-squares \citep[OLS(Y$|$X),][]{1990ApJ...364..104I}, bisector regression \citep{1990ApJ...364..104I}, Bayesian linear-regression \citep{2013PASP..125..306F}, Theil-Sen \citep{Sen01121968}, orthogonal distance regression \citep[ODR,][]{Boggs1992ODRPACK}, bivariate correlated errors and intrinsic scatter estimators \citep[BCES(Y$|$X),][]{1996ApJ...470..706A}, York \citep{2004AmJPh..72..367Y}, linmix \citep{2007ApJ...665.1489K}, and Hyper-Fit \citep{2015PASA...32...33R}. We then evaluated each method according to two criteria that are relevant for this work, which are the scatter around the relation and the strength of any remaining residual trend with optical diameter. The scatter is measured as the standard deviation of the residuals from the best-fit line ($\sigma_{\rm obs} = \mathrm{std}(\Delta\log D_{\rm HI})$). The strength of the residual trend with optical diameter is estimated by calculating $\Delta \log D_{\rm HI} = \log D_{\rm HI} - (\alpha\,\log D_{25} + \beta)$ for each galaxy and computing the following statistical parameters: 
\begin{enumerate}
  \item the Spearman rank-correlation coefficient $\rho_{\rm S}$
    between the residuals and $\log D_{25}$, and its
    two-sided $p$-value $p_{\rm S}$;
  \item the mean residual in the subsample of galaxies whose
    $\log D_{25}$ is below the sample median,
    $\langle \Delta \rangle_{\rm low}$;
  \item the mean residual in the subsample of galaxies whose $\log D_{25}$ is above the sample median, $\langle \Delta \rangle_{\rm high}$. 
\end{enumerate}
\begin{table*}
\centering
\caption{Comparison of different fitting methods}
\begin{tabular}{lccccccc}
\toprule \toprule
Method & $\alpha$ & $\beta$ & $\sigma_{\rm obs}$ & $\rho_{\rm S}$ & $p_{\rm S}$ & $\langle \Delta \rangle_{\rm low}$ & $\langle \Delta \rangle_{\rm high}$ \\
\midrule
\multicolumn{8}{l}{Consistent with zero $D_{25}$ trend}\\
Theil--Sen & $0.719$ & $+0.667$ & $0.153$ & $-0.006$ & $0.906$ & $+0.005$ & $+0.002$ \\
Bayesian (adopted) & $0.704$ & $+0.691$ & $0.153$ & $+0.014$ & $0.781$ & $-0.002$ & $+0.000$ \\
OLS($Y|X$) & $0.696$ & $+0.701$ & $0.153$ & $+0.025$ & $0.62$ & $-0.002$ & $+0.002$ \\
\texttt{linmix} & $0.736$ & $+0.648$ & $0.153$ & $-0.028$ & $0.579$ & $+0.004$ & $-0.004$ \\
BCES($Y|X$) & $0.738$ & $+0.645$ & $0.153$ & $-0.031$ & $0.535$ & $+0.005$ & $-0.005$ \\
\midrule
\multicolumn{8}{l}{Noticeable trend with $D_{25}$; not adopted}\\
Bisector & $0.998$ & $+0.305$ & $0.166$ & $-0.346$ & $7.2\times 10^{-13}$ & $+0.047$ & $-0.048$ \\
\textsc{hyper-fit} & $1.031$ & $+0.261$ & $0.169$ & $-0.380$ & $2.0\times 10^{-15}$ & $+0.054$ & $-0.052$ \\
York & $1.219$ & $-0.039$ & $0.189$ & $-0.550$ & $1.4\times 10^{-33}$ & $+0.137$ & $-0.031$ \\
ODR & $1.219$ & $-0.039$ & $0.189$ & $-0.550$ & $1.4\times 10^{-33}$ & $+0.137$ & $-0.031$ \\
\bottomrule
\end{tabular}
\tablefoot{\label{table:trend_test}Testing different fitting methods to adopt for the baseline $\log D_{25}$--$\log D_{\rm HI}$ relation in order to assess whether the resulting residuals, $\Delta \log(D_{\rm HI})$, show any dependence on optical diameter. For each estimator we report the slope $\alpha$ and intercept $\beta$ of Eq.~\eqref{eq:sizesize}, the observed residual scatter $\sigma_{\rm obs}$ (in dex), the residual Spearman rank-correlation coefficient $\rho_{\rm S}$ with $\log D_{25}$ and its two-sided $p$-value $p_{\rm S}$, and the mean residuals in the low- and high-$D_{25}$ halves of the sample as explained in the text, $\langle \Delta \rangle_{\rm low}$ and $\langle \Delta \rangle_{\rm high}$ (in dex). The estimators are grouped by whether their residuals are consistent with zero $D_{25}$ trend (top block) or show a noticeable trend (bottom block), and sorted within each block by $|\rho_{\rm S}|$. The adopted baseline estimator is the Bayesian (\texttt{emcee}) fit.}
\end{table*}
\begin{figure*}
\centering
\includegraphics[width=\textwidth]{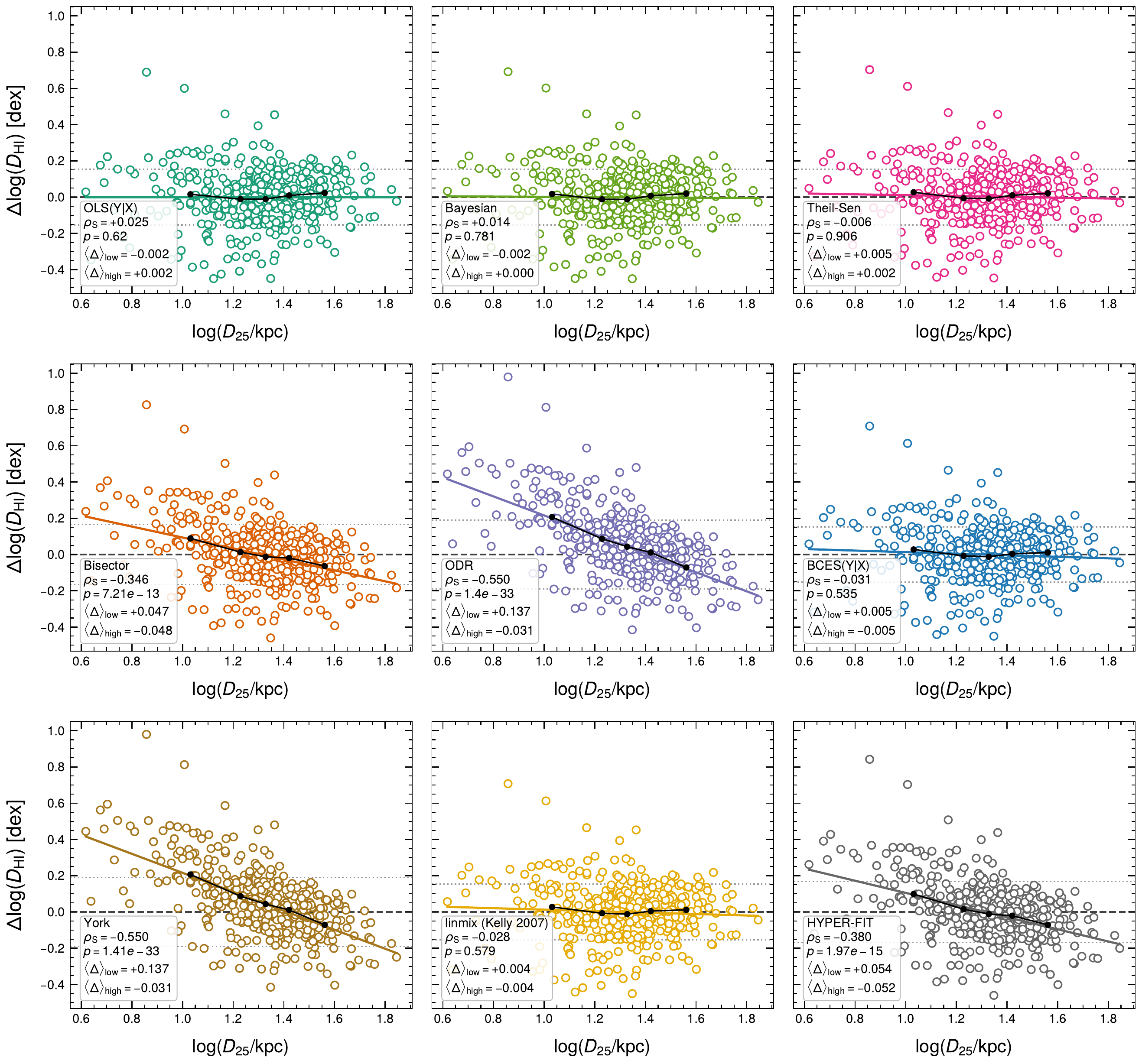}
\caption{\label{fig:trend_test}The $\Delta \log D_{\rm HI}$ vs $\log D_{25}$ relation of the combined AMIGA sample fitted using the nine linear estimators mentioned in Sect.~\ref{sub:diameter-optical}. The solid lines are the fits to the data points, the black dots are quintile-binned medians of the residuals, and the horizontal dotted lines indicate $\pm \sigma_{\rm obs}$.}
\end{figure*}
We considered an estimator as free of $D_{25}$-dependent residual bias when $\rho_{\rm S}$ is statistically indistinguishable from zero, $p_{\rm S}$ exceeds a conventional significance threshold, and the mean residuals in both the low- and high-$D_{25}$ bins are each consistent with zero. The results of our test are summarized in Table~\ref{table:trend_test} and presented in Fig.~\ref{fig:trend_test}. The test indicates that the bisector, ODR, York, and \textsc{hyper-fit} fits resulted in a noticeable size-dependent bias. However, OLS($Y|X$), Theil--Sen, BCES($Y|X$), \texttt{linmix}, and the Bayesian \texttt{emcee} fits are statistically indistinguishable and consistent with no size-dependent bias. Thus, we adopt the Bayesian \texttt{emcee} fit as our preferred method to fit the AMIGA baseline. The Bayesian \texttt{emcee} fitted to the AMIGA galaxies yields the relation
\begin{equation}\label{eq:baseline}
\log(D_{\rm HI}) = 0.704^{+0.037}_{-0.037}\,\log(D_{25})
                 + 0.691^{+0.049}_{-0.050},
\end{equation}
corresponding to a sub-linear scaling $D_{\rm HI} \propto D_{25}^{0.70}$, with a Pearson correlation coefficient $r = 0.70$ ($r^{2} = 0.49$) and an observed residual scatter $\sigma_{\rm obs} = 0.15$~dex. The best-fit line and the $\pm3\sigma$ scatter band are overlaid on the $D_{\rm HI}$--$D_{25}$ plane in Fig.~\ref{fig:diameter_correlation}.
Figure~\ref{fig:diameter_correlation} also presents the residuals as a function of $\log(D_{25})$. No trend with optical size is apparent in either sample, confirming that the offset between the two populations is not driven by their different $D_{25}$ distributions. About \HCGBelowBaselinePercent\% of HCG galaxies with a measured \HI\ diameter lie below the AMIGA fitted baseline. Including the \HCGUpperLimitCount\ members with upper-limit \HI\ diameters (nondetections plus three beam-limited members) as upper limits raises the fraction below the baseline to $\sim$\HCGKMBelowBaselinePercent\% (Kaplan--Meier). \\ \indent Our AMIGA calibration is noticeably shallower than the $D_{\rm HI}$--$D_{25}$ relations of \HI-selected compilations. Applying the same Bayesian estimator to the 108 field spirals and irregulars of \citet{B1997} and to the \citet{2016MNRAS.460.2143W} compilation yields nearly linear scalings, $D_{\rm HI} \propto D_{25}^{0.94\pm0.03}$ and $D_{\rm HI} \propto D_{25}^{0.92\pm0.03}$, respectively, over the same $D_{25}$ range spanned by AMIGA. We attribute this difference primarily to the sample selection criteria rather than to the fitting. Interferometric \HI\ compilations are effectively \HI-selected and therefore favor gas-rich galaxies with extended \HI\ disks, whereas AMIGA is an optically selected sample defined solely by isolation, so at fixed $D_{25}$ it samples the full range of \HI\ extents of secularly evolving galaxies. We deliberately do not adopt these larger literature calibrations as our baseline as they mix galaxies from all environments, including cluster members whose \HI\ disks are already truncated, so using them as the reference would absorb part of the environmental signal we aim to measure. With 407 galaxies, the AMIGA fit is not limited by statistics. We also tested whether the relation prefers different slopes in different diameter ranges by fitting independent lines to the galaxies below and above the median $D_{25}$; the two-segment fit does not improve on the single power law (F-test $p=0.20$, $\Delta\mathrm{BIC}=+9$ in favor of the single line), so we adopt a single slope over the full range.
Figure~\ref{fig:residuals_hist} shows the density-normalized distributions
of the residuals. The AMIGA histogram is broadly distributed and
centered at zero, as expected from the reference baseline, with comparable
weight on both the positive and negative sides. The HCG distribution is strongly skewed toward small \HI\ diameters and remarkably non-Gaussian. The skewness is mostly driven by galaxies from Phase 3. Because more than half of the HCG members are \HI\ nondetections, the Kaplan--Meier median of the full sample lies in the censored tail, below the peak formed by the detections alone; the value marked in Fig.~\ref{fig:residuals_hist} is its rigorous upper bound (for comparison, the detections-only median is $-0.10$~dex). The median offset implies that, typically, HCG galaxies
have \HI\ disks about $\sim$\HCGKMSizeFractionPercent\% of the size expected for
isolated galaxies of the same optical diameter.

For galaxies with measured $D_{\rm HI}$, we test whether the AMIGA and HCG residuals could plausibly be drawn from
the same parent distribution by applying complementary two-sample tests
that make minimal assumptions about the data (Mann--Whitney $U$,
Kolmogorov--Smirnov, and Anderson--Darling). All three reject the null
at very high significance, as summarized in Table~\ref{table:stat_tests}. Including the upper limits, the Gehan generalized Wilcoxon test (which accounts for the censoring) likewise rejects the null at very high significance ($p=\GehanP$; Table~\ref{table:stat_tests_censored}). Our Cliff's $\delta$ effect-size measure indicates that a randomly chosen AMIGA galaxy has a larger residual than a
randomly chosen HCG galaxy roughly \CliffProbAmigaGreaterHCGCensoredPercent\% of the time.
\begin{table}
\centering
\caption{\label{table:stat_tests}Statistical tests comparing the AMIGA and HCG residual distributions.}
\resizebox{0.47\textwidth}{!}{%
\begin{tabular}{lcc}
\toprule \toprule
Test & Statistic & $p$-value \\
\midrule
Mann--Whitney $U$         & $14149$   & $2.33 \times 10^{-5}$ \\
Kolmogorov--Smirnov       & $D = 0.323$ & $1.03 \times 10^{-4}$ \\
Anderson--Darling          & $12.093$ & $< 0.001$ \\
\midrule
\multicolumn{3}{l}{\textit{Effect sizes}} \\
\midrule
Cliff's $\delta$          & \multicolumn{2}{c}{$+0.363$ ($P(\mathrm{AMIGA} > \mathrm{HCG}) = 0.682$)} \\
\bottomrule
\end{tabular}%
}
\tablefoot{}
\end{table}

\begin{table}
\centering
\caption{\label{table:stat_tests_censored}Survival-analysis comparison of the AMIGA and HCG residual distributions.}
\begin{tabular}{lcc}
\toprule \toprule
Statistic & AMIGA & HCG (det+lim) \\
\midrule
$N$ & 407 & 124 (73 lim) \\
KM median $\Delta$ [dex] & $+0.005$ & $+nan$ \\
\% below baseline & 49\% & 90\% \\
\midrule
\multicolumn{3}{l}{Gehan generalised Wilcoxon (censored):} \\
\multicolumn{3}{l}{\quad $z=-11.91$, $p=1.1\times10^{-32}$} \\
\bottomrule
\end{tabular}
\tablefoot{The \HI\ nondetected HCG members (Table~\ref{table:upperlimits}) are included as beam-size upper limits (left-censored data). Medians and the fraction below the baseline are Kaplan--Meier estimates; the two-sample comparison uses the Gehan generalized Wilcoxon test. The HCG sample here is the full \HCGSampleSizeCensored\ members (\HCGUpperLimitCount\ upper limits); the AMIGA sample is unchanged.}
\end{table}

\begin{figure*}
    \sidecaption
     \includegraphics[width=12cm]{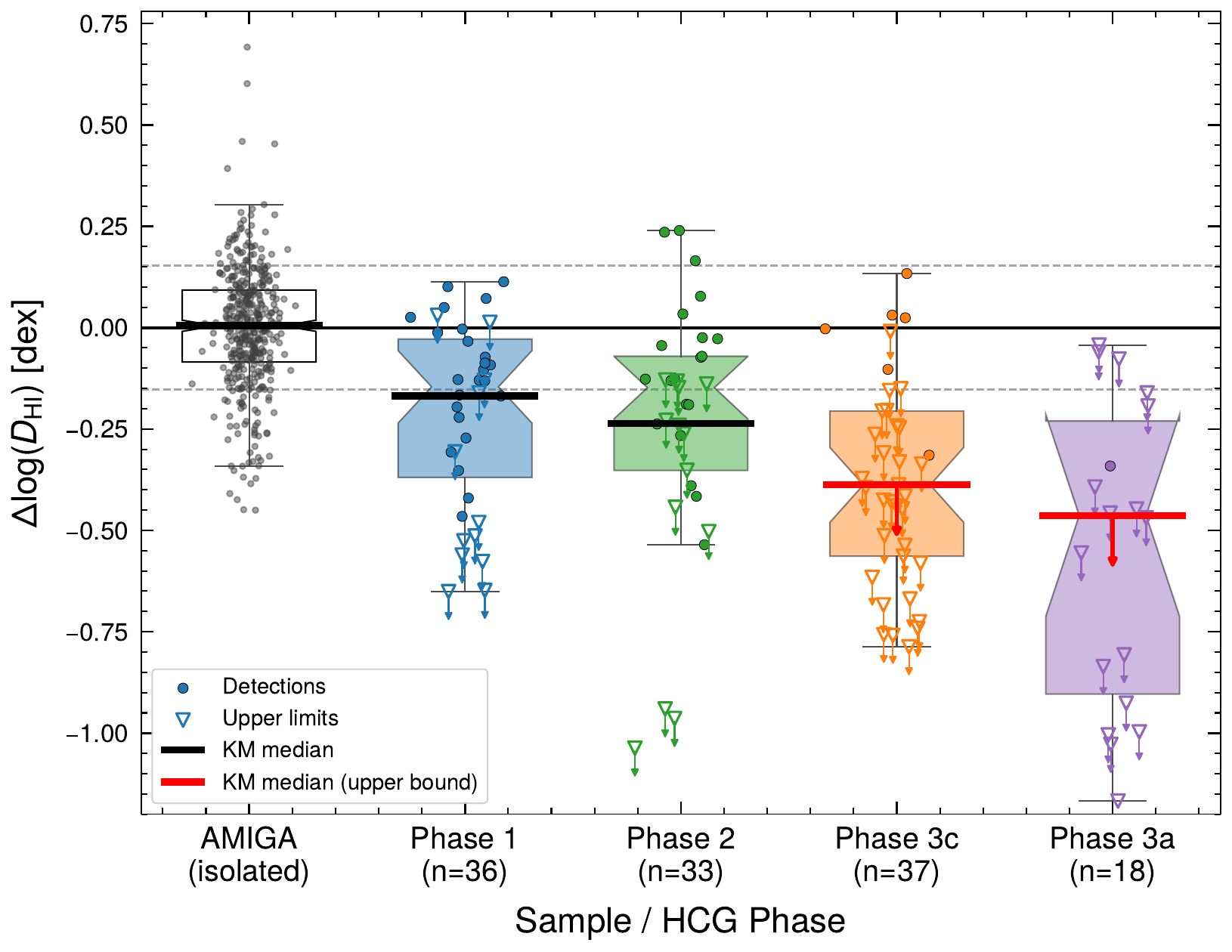}
	\caption{Box-and-whisker distributions of the residuals $\Delta\log(D_{\rm HI})$ for the AMIGA reference and the HCG galaxies grouped by evolutionary phase. Each box spans the interquartile range of the plotted values, the notch marks the confidence interval on the sample median, and the whiskers extend to the most extreme points within 1.5 times the interquartile range beyond the box; no outliers are drawn. Filled circles are detections and open downward triangles are beam-size upper limits; the boxes are drawn with the upper limits placed at their limit value (a conservative view, since the true diameters are smaller). Bold bars mark the Kaplan--Meier median (black, defined for AMIGA and Phases~1--2) or its upper bound (red with downward arrow, for Phases~3c and 3a where more than half the members are upper limits). The horizontal solid line marks $\Delta = 0$ and the dashed lines indicate the $\pm 1\sigma_{\rm obs}$ band of the AMIGA baseline scatter ($\sigma_{\rm obs} = 0.153$~dex).}
    \label{fig:hist_by_phase}
\end{figure*}

Figure~\ref{fig:hist_by_phase} presents the residual distributions broken down by HCG evolutionary phase. In each phase, the median sits below the AMIGA baseline. From Phase 1 to Phase 3, the $\Delta\log(D_{\rm HI})$ distribution progressively shifts toward galaxies with \HI\ diameters further below the AMIGA expectation, with the strongest deviation seen in the most evolved phase, where the population is largely composed of galaxies with severely or completely depleted \HI\ disks that fall below the detection limit. For ease of interpretation, we convert the residuals $\Delta \equiv \Delta\log(D_{\rm HI})$ into an \HI\ diameter deficit of a galaxy relative to the AMIGA expectation at fixed $D_{25}$,
\begin{equation}\label{eq:truncation_indices}
\mathrm{Deficit}\,[\%] \;=\; 100 \times \left(1 - \frac{D_{\rm HI}}{D_{\rm HI,\,exp}(D_{25})}\right) \;=\; 100 \times \left(1 - 10^{\Delta}\right),
\end{equation}
where $D_{\rm HI,\,exp}(D_{25})$ is the AMIGA-baseline prediction at the galaxy's optical diameter. A deficit of $0\%$ indicates an \HI\ disk identical in size to that of an isolated galaxy of the same optical diameter; a deficit of $50\%$ indicates an \HI\ disk half as large; negative deficits would indicate enhancement rather than truncation. We summarize the median values of $\Delta\log(D_{\rm HI})$ and the corresponding median \HI\ diameter deficit for each phase in Table~\ref{table:phase_stats}.
\begin{table}
\centering
\caption{\label{table:phase_stats}Residual statistics by HCG evolutionary phase.}
\begin{tabular}{lcccc}
\toprule
Phase & $n$ & Mean $\Delta$ & Median $\Delta$ & Median $D_{\rm{HI}}$  \\
      &     &               &                 & deficit \\
      &     & [dex]         & [dex]           & [\%]    \\
\midrule
1     & 24  & $-0.125$      & $-0.116$        & 23 \\
2     & 20  & $-0.105$      & $-0.099$        & 20 \\
3c     & 6   & $-0.039$      & $+0.011$        & -2 \\
\bottomrule
\end{tabular}
\tablefoot{Mean and median values of $ \Delta\log(D_{\rm HI})$ for each HCG phase. The number of galaxies considered is presented in the second column. The corresponding \HI\ diameter deficits (see Eq.~\ref{eq:truncation_indices}) are shown in the last column.}
\end{table}

\begin{table}
\centering
\caption{\label{table:phase_stats_censored}Residual statistics by HCG evolutionary phase, including the upper limits.}
\begin{tabular}{lccccc}
\toprule \toprule
Phase & $N$ & $N_{\rm lim}$ & KM median $\Delta$ & \% below & KM deficit \\
      &      &              & [dex]               & baseline & [\%] \\
\midrule
1 & 36 & 12 & $-0.169$ & 86\% & 32 \\
2 & 33 & 13 & $-0.238$ & 85\% & 42 \\
3c & 37 & 31 & $\leq -0.39$ & 92\% & $\geq 59$ \\
3a & 18 & 17 & $\leq -0.46$ & 100\% & $\geq 66$ \\
\bottomrule
\end{tabular}
\tablefoot{The \HI\ nondetected members are included as beam-size upper limits (Table~\ref{table:upperlimits}). Medians and the fraction below the baseline are Kaplan--Meier estimates; $N_{\rm lim}$ is the number of upper limits in each phase. For Phase~3c and Phase~3a more than half of the members are upper limits, so the Kaplan--Meier median is unconstrained; we instead quote a rigorous upper bound (the median of the per-galaxy upper limits), hence the ``$\leq$'' / ``$\geq$'' entries.}
\end{table}

\subsection{\HI\ disk truncations across different environments}\label{sec:survey_comparison}

\begin{figure*}
    \centering
    \begin{tabular}{c c}
       \includegraphics[scale=0.4]{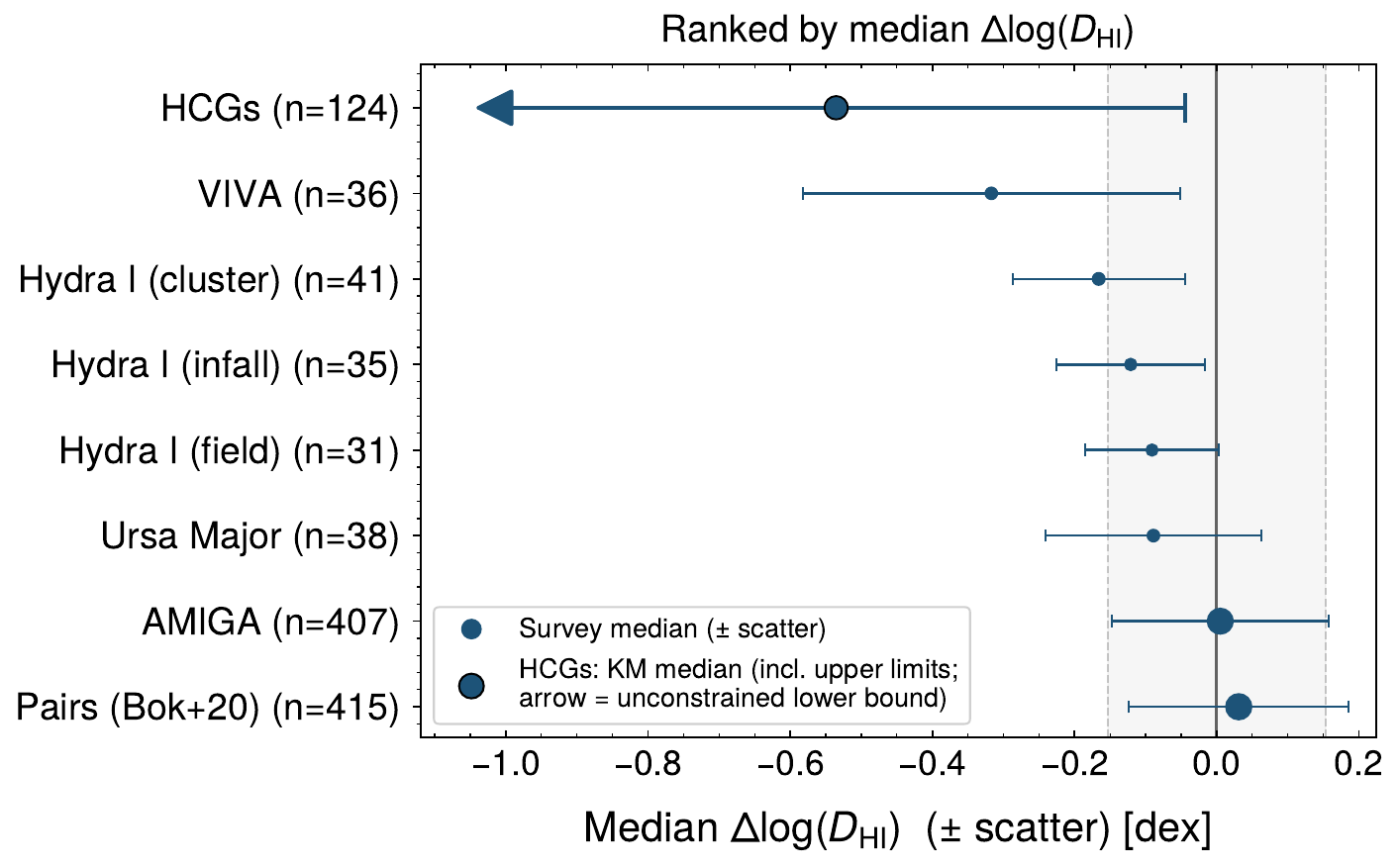}&
       \includegraphics[scale=0.5]{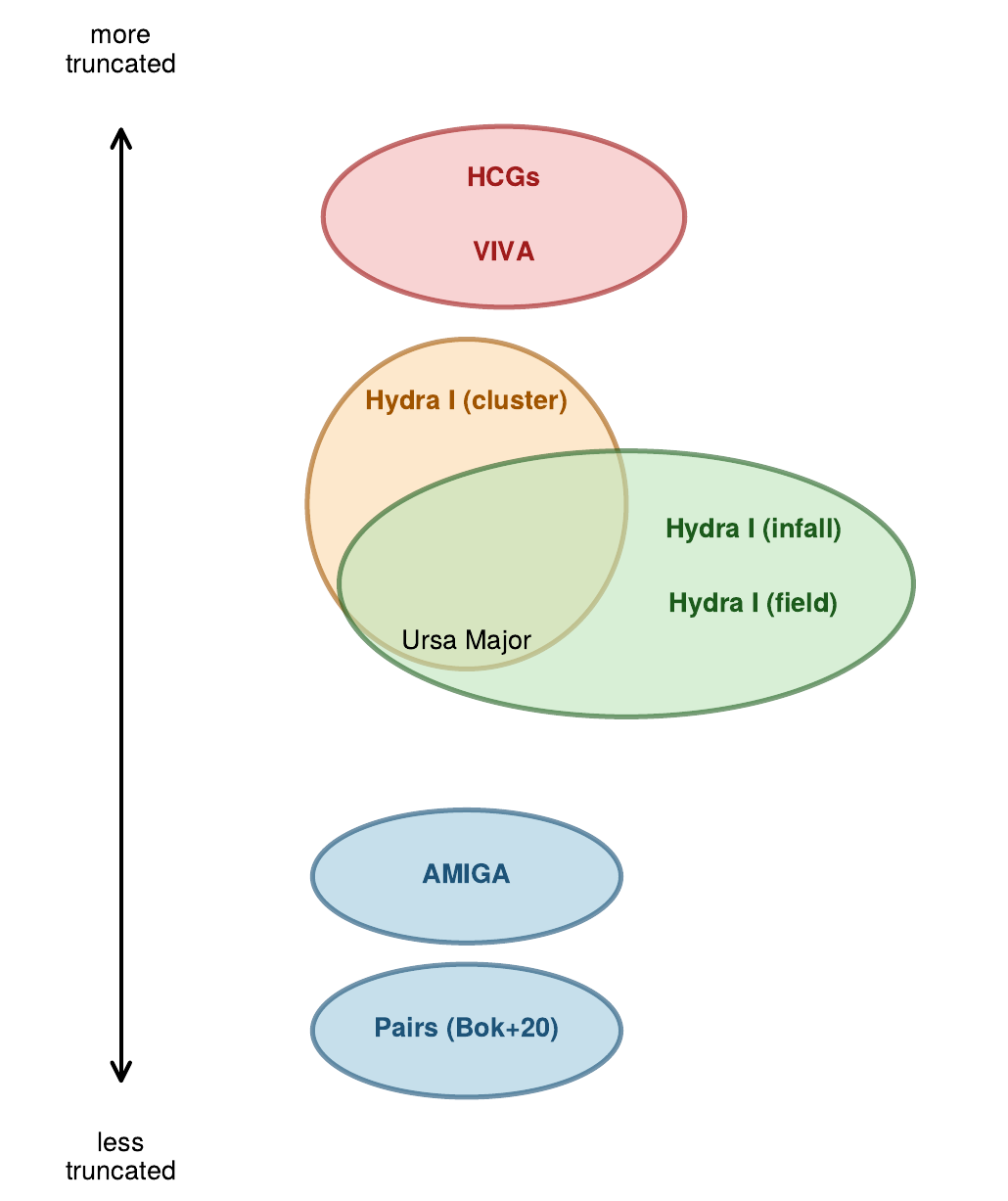}
       \end{tabular}
	\caption{Left panel: Median residual $\Delta\log(D_{\rm HI})$ relative to the AMIGA baseline for each survey. The error bars indicate the $\pm 1\sigma$ scatter around the median. Surveys are ranked from most truncated (top) to least truncated (bottom). The vertical solid line marks $\Delta = 0$ (the AMIGA expectation), and the shaded region indicates the $\pm 1\sigma_{\rm AMIGA}$ envelope. The marker size is proportional to the sample size. The HCG point is the Kaplan--Meier median of the full detection-plus-upper-limit sample ($N=124$); its bar reaches the KM 84th percentile on the upper (less-truncated) side, while the lower side is unconstrained (arrow), since more than half of the HCG members are upper limits. Right panel: A diagram illustrating the statistical significance of the difference between the median residual $\Delta\log(D_{\rm HI})$ of each sample. Samples enclosed by the same ellipse are statistically indistinguishable. Samples that do not share an ellipse are statistically different. The vertical ordering indicates increasing H\,{\sc i} truncation relative to the AMIGA baseline at the bottom.}
       \label{fig:survey_median}
\end{figure*}

To benchmark the HCG sample against other \HI\ surveys with well-defined environments, we applied the same AMIGA baseline (Eq.~\ref{eq:baseline}) to four additional samples, namely the VIVA \citep[Virgo cluster;][]{2009AJ....138.1741C} and Ursa Major (loose group) surveys compiled in \citet{2016MNRAS.460.2143W}, the Hydra I (cluster, infall and field) survey by \citet{2022MNRAS.510.1716R}, and the galaxy-pair sample from the ALFALFA $\alpha.70$ catalog studied by \citet{2020MNRAS.499.3193B}. Below is a brief description of each additional survey used. 
\begin{enumerate}
	\item Bok+20 Pairs of galaxies: \citet{2020MNRAS.499.3193B} selected a sample of mostly spiral+spiral pairs of galaxies from the ALFALFA $\alpha.70$ catalog \citep{2018ApJ...861...49H}. Their sample was selected to have a projected separation less than 100 kpc, detected \HI\ masses above $10^9\,M_\odot$ for both members, and a line-of-sight velocity difference of less than 1000 $\mathrm{km~s^{-1}}$ between the two galaxies. Their sample includes isolated pairs, triples, and compact group candidates. The \citet{2020MNRAS.499.3193B} catalog does not have resolved \HI\ diameters, so we infer their $D_{\mathrm{HI}}$ from the ALFALFA \HI\ mass with the same Bayesian size--mass relation used for the AMIGA single-dish detections. Because these diameters are inferred from the \HI\ mass, the comparison for this sample essentially probes an \HI-mass deficiency expressed in diameter units rather than a directly measured $D_{\rm HI}$ distribution. Through the adopted size--mass relation, the median offset we measure for the pairs ($\Delta\log D_{\rm HI} = +0.03$~dex; Table~\ref{table:survey_residuals}) corresponds to a median \HI\ mass ${\sim}0.06$~dex above the isolated expectation, that is, the pairs are not \HI\ deficient. This agrees with previous \HI-content studies of pairs: \citet{2020MNRAS.499.3193B} found that these same pairs differ from isolated galaxies mainly through a larger dispersion in \HI\ deficiency rather than a systematic offset, \citet{2017MNRAS.466.4795J} reported enhanced \HI\ fractions in the smallest groups, and interacting pairs are generally observed to retain their \HI, redistributed into tidal features rather than removed \citep{1996AJ....111..655H}, with gas fractions that remain substantial and fuel their star formation \citep{2015MNRAS.449.3719S}. We obtain their $D_{25}$ from HyperLEDA \citep{2014A&A...570A..13M}. \\
	\item Ursa Major: This sample consists of 38 spiral galaxies with measured $D_{\mathrm{HI}}$ observed by \citet{2001A&A...370..765V} using the Westerbork Synthesis Radio Telescope (WSRT). As reported by \citet{2001A&A...370..765V}, the Ursa Major cluster has very different properties than more massive and denser clusters like Virgo. It is dominated by late type galaxies, it does not have a dense concentration of galaxies in its core, it has a low velocity dispersion ($\sigma_{v}~158~km~s^{-1}$), it is dynamically cold with no detected X-ray-emitting intracluster medium \citep[see also][]{2014MNRAS.445..630P}. Overall, its environmental conditions are in many respects similar to those of the field. We therefore treat Ursa Major as a field-like, low-density cluster reference sample rather than as a benchmark for strong environmentally driven processing. \\
  \item Hydra I (Abell 1060): Hydra\,I presents a mixed environmental state. It has long been regarded as an evolved, dynamically relaxed cluster based on its regular X-ray morphology and its early-type dominated core \citep{1980ApJ...242..469W, 2010A&A...520L...9V,2022A&A...659A..92L, 2022A&A...668A.184H}. However, it is also relatively spiral-rich, contains a number of gas-rich dwarfs \citep{1999A&AS..136..539D}, and includes central spiral galaxies that are not strongly H\,{\sc i} deficient \citep{1983A&A...125..187R}. These properties indicate that, although the cluster core is dynamically mature, a substantial fraction of the late-type population has not yet been fully stripped. This is supported by the WALLABY analysis of \citet{2021ApJ...915...70W}, who found that many H\,{\sc i}-detected Hydra galaxies are likely undergoing early-stage ram-pressure stripping. A recent study by \citet{2025A&A...704A.264D} reinforces the composite state of Hydra I. Using DECam $ugriz$, H$\alpha$ imaging, and MeerKAT \HI\ observations, they find an evolved cluster population alongside recently accreted, gas-rich galaxies with disturbed \HI\ morphologies and enhanced star formation. \\
  \item VIVA (Virgo cluster): The 36 VIVA galaxies we use in this study are drawn from the VLA imaging of Virgo spirals by \citet{2009AJ....138.1741C} and were used in the size--mass study by \citet{2016MNRAS.460.2143W}. Out of the 36 VIVA galaxies, 30 are spirals, two have irregular Magellanic types, three are lenticulars, and one is a dwarf elliptical \citep{2009AJ....138.1741C}. The full VIVA survey contains 53 galaxies spanning projected distances of $d_{87} = 0.3$--$3.3$~Mpc from M87, sampling low- to high-density regions of Virgo; the 36 galaxies used here are the subset retained by \citet{2016MNRAS.460.2143W}, who required the interferometric and single-dish \HI\ fluxes to agree within 15\%. This subset preserves the radial coverage of the parent survey ($d_{87} = 0.4$--$3.4$~Mpc, median $1.1$~Mpc), extending from the cluster core, defined by \citet{2009AJ....138.1741C} as $d_{87} \lesssim 0.5$~Mpc, out to beyond the virial radius. Virgo has been classified as a dynamically young, unrelaxed cluster, as characterized by its different substructures and the infall of late-type galaxies \citep{1985AJ.....90.1681B, 2001ApJ...559..791C, 2006PASP..118..517B}. It is similar to the Hydra I cluster in terms of virial mass and velocity dispersion \citep[$\sigma_{v}~638~km~s^{-1}$, see][]{2020A&A...635A.135K}. Its core consists of mainly early-type gas-deficient galaxies \citep{2020A&A...635A.135K}. It has hot intracluster gas, causing morphological transformation of galaxies as they fall toward the cluster center \citep{2016A&A...596A.101P}.  
\end{enumerate}	
\begin{table*}
\centering
\caption{\label{table:survey_residuals}Residual statistics $\Delta\log(D_{\rm HI})$ relative to the AMIGA baseline for various \HI\ surveys.}
\begin{tabular}{lcccccc}
\toprule \toprule
Sample & $\mathrm{N_{gal}}$ & Median $\Delta$ & $\sigma$ & $f_{\Delta < -\sigma}$ & $f_{\Delta > +\sigma}$ & Ref \\
 & & [dex] & [dex] & [\%] & [\%] & \\
\midrule
Pairs (Bok+20)     & 415 & $+0.031$ & 0.155 &  10.6 &  18.8 & \citet{2020MNRAS.499.3193B} \\
AMIGA              & 407 & $+0.005$ & 0.153 &  12.8 &  12.0 & this work \\
Ursa Major         &  38 & $-0.089$ & 0.152 &  42.1 &   5.3 & \citet{2001A-A...370..765V} \\
Hydra I (field)    &  31 & $-0.091$ & 0.094 &  16.1 &   0.0 & \citet{2022MNRAS.510.1716R} \\
HCGs               &  51 & $-0.103$ & 0.178 &  35.3 &   5.9 & this work \\
Hydra I (combined) & 107 & $-0.120$ & 0.114 &  37.4 &   0.9 & \citet{2022MNRAS.510.1716R} \\
Hydra I (infall)   &  35 & $-0.121$ & 0.104 &  37.1 &   2.9 & \citet{2022MNRAS.510.1716R} \\
Hydra I (cluster)  &  41 & $-0.166$ & 0.121 &  53.7 &   0.0 & \citet{2022MNRAS.510.1716R} \\
VIVA               &  36 & $-0.317$ & 0.265 &  75.0 &   2.8 & \citet{2009AJ....138.1741C} \\
\bottomrule
\end{tabular}
\tablefoot{Columns: (1) The sample, (2) Number of galaxies, (3) Median residual offset from the AMIGA baseline, (4) Standard deviation of the residuals, (5) Fraction of galaxies with $\Delta\log(D_{\rm HI}) < -\sigma_{\rm AMIGA}$, (6) Fraction of galaxies with $\Delta\log(D_{\rm HI}) > +\sigma_{\rm AMIGA}$, (7) References. $^{\dagger}$For the HCGs, columns (2), (3), (5), and (6) include the \HCGUpperLimitCount\ \HI\ nondetected members as beam-size upper limits: the median and fractions are Kaplan--Meier estimates, and a single residual scatter (col.~4) is not quoted because of the censoring. All other surveys use detection-based statistics.}
\end{table*}
For each survey, we computed the residuals $\Delta\log(D_{\rm HI})$ relative to the AMIGA fit and derived the median offset, the scatter, and the fractions of galaxies having $\Delta\log(D_{\rm HI}) < -\sigma_{\rm AMIGA}$ (i.e., $< -\AmigaSigmaDex$~dex), and $\Delta\log(D_{\rm HI}) > +\sigma_{\rm AMIGA}$ (i.e., $> +\AmigaSigmaDex$~dex), where $\sigma_{\rm AMIGA}$ is the intrinsic scatter of the AMIGA baseline. The results are summarized in Table~\ref{table:survey_residuals}. Figure~\ref{fig:survey_median} shows the median $\Delta\log(D_{\rm HI})$ and scatter for each survey, ordered by increasing offset from the AMIGA baseline. The median residuals span $\sim0.6$~dex relative to the AMIGA reference. To test whether these differences are statistically significant, we performed pairwise comparisons between the residual distributions. For comparisons involving upper limits, we used the Gehan generalized Wilcoxon test; all other pairs were compared with two-sided Mann--Whitney $U$ tests. In both cases, we applied a Benjamini--Hochberg false-discovery-rate correction at $\alpha=0.05$ \citep{benjaminitest}, supplemented by bootstrap confidence intervals \citep{Efron1979Bootstrap}. Figure~\ref{fig:survey_median} presents a diagram of the statistical significance of the difference between the median residual $\Delta\log(D_{\rm HI})$ of each sample. Samples enclosed by the same ellipse are statistically indistinguishable, while those that do not share an ellipse are statistically different. HCGs have the most negative median residual and are statistically indistinguishable from the VIVA sample. The truncation in this sample is not confined to the cluster core: the residuals show only a weak dependence on clustercentric distance (Spearman $\rho = +0.28$, $p = 0.10$), and strongly truncated members are found out to $d_{87} \sim 2$~Mpc, consistent with the truncated \HI\ disks that \citet{2009AJ....138.1741C} report at large clustercentric distances. They are followed by Hydra I (cluster), Hydra I (infall), Hydra I (field), and Ursa Major. Ursa Major is statistically indistinguishable from both Hydra I (cluster) and the Hydra I infall/field subsets. Hydra I (infall) and Hydra I (field) are statistically indistinguishable from each other and from Ursa Major, but they are offset from the AMIGA reference. At the least-truncated end sits AMIGA and the Bok+20 pairs, which are statistically different from each other and are significantly offset from all other samples. It is not surprising to see the Bok+20 pairs have statistically larger \HI\ disks than the AMIGA. As reported by \citet{2017MNRAS.466.4795J}, galaxies in small groups of N=2 members can have higher \HI\ fraction than isolated galaxies at the same stellar mass. In addition, the Bok+20 sample was selected such that both pair members have $M_{\mathrm{HI}}>10^{9}M_{\odot}$. Thus, stripped galaxies or \HI-poor companions are not included in their catalogs. As a result, the difference is also partly due to a selection bias.        
\section{Discussion}\label{discussion}
Given the tightness of the \HI\ size--mass relation, the prevalence of truncated \HI\ disks in rich environments is expected; galaxies in clusters and groups are well known to be \HI\ deficient \citep[e.g.,][]{2006PASP..118..517B, 2021PASA...38...35C, 2022A&ARv..30....3B}, and a reduced \HI\ mass at a roughly universal mean \HI\ surface density implies a smaller \HI\ diameter. What our residual-based analysis adds is a quantification of this effect at fixed optical size, and of how it develops along the compact-group evolutionary sequence.\\ \indent In Phase 1 (groups in which the \HI\ is still mostly confined to the galaxy disks), galaxies in HCGs appear to be more truncated relative to the AMIGA baseline. This suggests that the outer \HI\ has already been removed by the time a group reaches that phase. This is consistent with the simulations of \citet{2005MNRAS.357L..21B}, which show that tidal forces in a group potential can efficiently strip the low-density outer \HI\ disk of an orbiting galaxy. Phase 2 has the largest spread in $\Delta\log(D_{\rm HI})$ most likely because this phase mixes galaxies at different stripping stages within the same evolutionary bin. Within Phase~2 we find galaxies spanning a continuous range of truncation, from near the baseline down to ${\sim}30\%$ of the expected \HI\ diameter, with several members whose \HI\ disks are comparable to the beam and thus close to the detection limit. The disk is therefore likely stripped progressively rather than in a single step. Galaxies in Phase~3c appear statistically less truncated than those in Phase~3a. This difference is because Phase~3c systems are defined as groups in which at least one galaxy still retains a significant \HI\ reservoir, possibly indicating the recent accretion of a gas-rich member into an otherwise evolved group environment. Of the Phase 3c members, only six have resolved \HI\ disks. These have a median residual consistent with zero ($\Delta\log D_{\rm HI}=+0.010$~dex), with five of the six within $1\sigma_{\rm obs}$ of the AMIGA baseline, and are statistically indistinguishable from the isolated galaxies (Mann--Whitney $p=0.69$). In contrast, the Phase 1 and Phase 2 detections are significantly truncated relative to isolated galaxies ($p<0.001$ and $p=0.013$). The detected Phase 3c galaxies are therefore consistent with relatively unperturbed, isolated-like systems, supporting the interpretation that the gas-rich member of a Phase 3c group is a recently accreted galaxy rather than a long-term survivor. However, we want to stress that this result is based on only the six detections we have and while consistent with the isolated baseline, the Phase 3c detections are not statistically distinguishable from the Phase 1 distribution. The measured HCG offset in Phase~3c represents a mixture of evolved group members and relatively unprocessed, recently accreted galaxies, thereby weakening the overall \HI\ truncation signal. In contrast, Phase~3a systems represent a more advanced evolutionary stage, where the group members have been severely or completely stripped of their \HI\ content. Their \HI\ disks are significantly more truncated relative to the AMIGA reference. \\ \indent  
The outer \HI\ disk is more sensitive to environmental processing than the optical disk. Our analysis shows that HCG galaxies have a median \HI\ diameter deficit of at least $\sim$\HCGKMDeficitPercent\% relative to isolated galaxies of the same optical diameter (Kaplan--Meier median residual $\lesssim\HCGKMMedianDex$~dex). These results indicate that environmental processes preferentially strip the low-density outer gas, while the inner stellar disk is affected far less. The stellar disk is nonetheless not entirely unaffected. \citet{2013MNRAS.434..325F} found that isolated late-type AMIGA galaxies have stellar disks about $1.2$ times larger in effective radius than less isolated galaxies of the same stellar mass. 
The AMIGA sample has a larger number of galaxies with extended \HI\ disk compared to galaxies in groups and clusters. This suggests that extreme isolation not only prevents gas removal but may also allow galaxies to accumulate or preserve large reservoirs of atomic gas in their outer disks over cosmic timescales. The wide scatter in the AMIGA sample hints at a diversity of evolutionary histories among isolated galaxies, possibly influenced by properties such as halo mass, angular momentum (Sorgho et al. 2026, in prep), or cold-mode accretion rates. Using ultra-deep optical imaging of the AMIGA sample, \citet{2023A&A...677A.117S} detected faint signatures of past interactions in ${\sim}40\%$ of strictly isolated galaxies, including tidal streams (${\sim}20\%$) and perturbed or asymmetric stellar haloes (${\sim}25\%$), alongside a high fraction of down-bending (Type~II) disk breaks ($56\%$). This shows that completely undisturbed evolution is rare even among highly isolated galaxies. It will be interesting to compare our results with simulated galaxies. We reserve this for future work.
\\ \indent 
A potential concern when combining samples observed with different instruments is that the heterogeneous angular resolution and sensitivity could bias the inferred \HI\ diameters, and hence the truncation we measure. We assessed this directly within our resolved AMIGA and HCG samples, which span a wide range of linear resolution ($2$--$37$~kpc) and column-density sensitivity ($\sim6\times10^{18}$--$4\times10^{20}\,\mathrm{cm^{-2}}$). The residual $\Delta\log(D_{\rm HI})$ shows no significant correlation with either the linear resolution (Spearman $\rho=0.07$--$0.22$, $p\geq0.11$) or the column-density limit ($\rho=0.06$, $p=0.66$), confirming that our beam-corrected diameters are not set by resolution or depth. The only marginal trend, with resolution, is that coarser beams yield slightly less truncation through beam smearing, so any residual effect would make the measured truncation conservative rather than spurious. Among the literature samples, Hydra\,I is the most distant ($\sim$61~Mpc) and hence the most coarsely resolved ($30\arcsec\approx8.9$~kpc, versus $\sim$1.2~kpc for VIVA and $\sim$2.7~kpc for Ursa Major), but the same conservative bias direction applies and all samples reach the $1\,M_\odot\,\mathrm{pc^{-2}}$ surface-density level at which $D_{\rm HI}$ is defined. We therefore conclude that the comparison between heterogeneous datasets is unlikely to introduce a systematic bias in our results.
\\ \indent 
The tightness of the HI mass--size relation across different environments reflects the near-universal shape of galaxy \HI\ surface density profiles, characterized by an outer exponential decline \citep{2014MNRAS.441.2159W, 2018AJ....155..233I, 2025ApJ...980...25W} and an inner region regulated by the conversion of atomic to molecular gas or stars \citep{2014MNRAS.441.2159W, 2019MNRAS.490...96S}. To move a galaxy off the relation may involve an extremely unlikely scenario, e.g., a gas disk being compressed without the formation of new stars.

\section{Summary}\label{summary}
We measured the \HI\ diameters of galaxies in HCGs and AMIGA using a direct ellipse fitting of the isophote at 1 $\mathrm{M_{\odot}\,pc^{-2}}$. We used the $\mathrm{D_{25}}$ isophotal diameter, measured at a surface brightness level of 25 B-mag arcsec\(^{-2}\), as a standard proxy for the optical diameters. We established the AMIGA isolated galaxy sample as a reference baseline in the $\log(D_{25})$--$\log(D_{\rm HI})$ plane. For each galaxy, we then measured the residual $\Delta\log(D_{\rm HI})$ as the difference between the observed \HI\ diameter and the \HI\ diameter predicted by the AMIGA relation at the same $D_{25}$. We define the sample offset as the median of these residuals such that negative values mean that the \HI\ disks are smaller than expected for isolated galaxies of the same optical size, while positive values mean that they are more extended. We also express this offset as an \HI\ diameter deficit, which measures the percentage by which $D_{\rm HI}$ is smaller than the AMIGA prediction at fixed $D_{25}$. We compared our measurements with literature samples. They include \HI-selected galaxy pairs from \citet{2020MNRAS.499.3193B}, the Ursa Major loose group \citep{2001A&A...370..765V}, the Virgo cluster (VIVA, \citealt{2009AJ....138.1741C}), and the Hydra\,I cluster split into cluster, infall, and field subsets \citep{2022MNRAS.510.1716R}.
Our main findings are summarized below.
\begin{itemize}[label=\textbullet]
	\item The AMIGA baseline follows $D_{\rm HI} \propto D_{25}^{0.70}$ with an observed residual scatter of $\sigma_{\rm obs} = \AmigaSigmaDex$~dex. The sublinear slope confirms that a fixed $D_{\rm HI}/D_{25}$ ratio cannot be used without introducing a size-dependent bias, and it is shallower than the nearly linear relations we derive for \HI-selected compilations with the same estimator (Sect.~\ref{sub:diameter-optical}).
	\item HCG galaxies have \HI\ disks that are at least $\sim$\HCGKMDeficitPercent\% smaller than expected for isolated galaxies of the same optical diameter, with $\sim$\HCGKMBelowBaselinePercent\% of members lying below the baseline.
	\item \HI\ truncation increases monotonically along the HCG evolutionary sequence. Phase~1 galaxies (members of groups whose \HI\ is still mostly confined to the galaxy disks) are already offset from the AMIGA baseline, indicating that truncation begins at an early stage, and becomes progressively stronger through Phase~2 to the most evolved phases (3c and 3a).
	\item The \HI-selected pairs of \citet{2020MNRAS.499.3193B} show no measurable outer \HI\ truncation. They occupy the least-truncated, baseline end of the distribution. Part of this behavior reflects their \HI-based selection, which requires both pair members to have $M_{\mathrm{HI}}>10^{9}\,M_{\odot}$ and therefore excludes strongly stripped systems.
	\item Comparison with the literature places HCGs at the most-truncated end of the environmental sequence. They are statistically indistinguishable only from the Virgo cluster sample (VIVA). The Hydra\,I cluster subset, its infall and field subsets, and the Ursa Major loose group occupy intermediate levels of truncation. Within Hydra\,I, the cluster subset exhibits stronger truncation than the infall and field regions.
\end{itemize}
\section{Data availability}
The data and analysis scripts required to fully reproduce the results, including the figures and tables in this work, are publicly available at \url{https://github.com/ianjarog/galaxydisksize}. The atlas of ellipse fits to the \HI\ iso-surface density contour for all galaxies in the AMIGA and HCG samples (three examples are shown in Fig.~\ref{fig:ellipse-fit}) is available as online material on Zenodo at \url{https://doi.org/10.5281/zenodo.21415323}.
\begin{acknowledgements}
RI, LV, BN, AS, PK, MK, SSE, JG acknowledge financial support from the grant
CEX2021-001131-S funded by MICIU/AEI/10.13039/501100011033. LV, BN, AS, RI, PK
acknowledge financial support from the grant PID2021-123930OB-C21 and
PID2024-155817OB-I00 funded by MICIU/AEI/10.13039/501100011033 and by ERDF/EU.
BN acknowledges financial support from the grant DGP\_POST\_2024\_01021 funded by
la Junta de Andaluc\'ia/CUII and by the ESF+. MK acknowledges support through the
SAFE -- ``Supporting At-Risk Researchers with Fellowships in Europe'' project,
funded by the European Union under the Grant Agreement ref.\ 101148426. Views and
opinions expressed are however those of the author(s) only and do not necessarily
reflect those of the European Union or the European Commission. RI, LV, BN, AS,
MK, SSE, JG acknowledge the Spanish Prototype of an SRC (SPSRC) service and
support funded by the Ministerio de Ciencia, Innovaci\'on y Universidades (MICIU),
by the Junta de Andaluc\'ia, by the European Regional Development Fund (ERDF) and
by the European Union NextGenerationEU/PRTR. The SPSRC acknowledges financial support from the Agencia Estatal de Investigaci\'on (AEI) through the ``Center of Excellence Severo Ochoa'' award to the Instituto de Astrof\'isica de Andaluc\'ia (IAA-CSIC) (SEV-2017-0709) \citep{10.1117/1.JATIS.8.1.011004}.
\end{acknowledgements}
\bibliographystyle{aa}
\bibliography{reference}
\nopagebreak
\begin{appendix}

\section{Fitting procedure}\label{app:fitting-procedure}
Following \citet{oy1998NumericallySD}, we describe the set of (x, y) points extracted at 1 $\mathrm{M_{\odot}~pc^{-2}}$ by the general conic equation:
\begin{equation}
    F(x,y) = ax^2 + 2bxy + cy^2 + 2dx + 2ey + f = 0,
\end{equation}
The parameters to fit are the coefficients $a,b,c,d,e,$ and $f$. The goal is to minimize the sum of the squared algebraic distances between the observed points and the conic: \[
\min \sum_{i=1}^{N} \left[ F(x_i, y_i) \right]^2
.\] To stabilize the solutions and ensure that $F(x,y)$ is an ellipse, the following constraint is set:
\begin{equation}
    ac-b^{2}=1.
\end{equation}
To rewrite the conic equation in a compact vector-matrix form, the following vectors are defined: 
\begin{equation}
\begin{aligned}
\mathbf{a} &= [\,a,\;b,\;c,\;d,\;e,\;f\,]^{\mathrm{T}} \\
\mathbf{x} &= [\,x^2,\; x y,\; y^2,\; x,\; y,\; 1\,]
\end{aligned}
\end{equation}
Thus, the problem can be reframed as finding the vector $\mathbf{a}$ that minimizes $||\mathbf{Da}||^2$. For optimal solutions, the following constraints are set: 
\begin{equation}\label{eq-constraint}
\begin{aligned}
    \mathbf{a}^\mathrm{T} \mathbf{C} \mathbf{a} = 1 \\
    \mathbf{S} \mathbf{a} = \lambda\, \mathbf{C} \mathbf{a},
\end{aligned}
\end{equation}
where the design matrix $\mathbf{D}$ is decomposed into linear and quadratic parts as follows. 
\begin{equation}
    \mathbf{D = (D_{1} | D_{2})},
\end{equation}
where

\begin{equation}
\mathbf{D_{1}} = 
\begin{pmatrix}
x_1^2 & x_1y_1 & y_1^2 \\
x_2^2 & x_2y_2 & y_2^2 \\
\vdots & \vdots & \vdots \\
x_n^2 & x_ny_n & y_n^2 
\end{pmatrix}
\qquad
\text{and}
\qquad
\mathbf{D_{2}} =
\begin{pmatrix}
x_1 & y_1 & 1 \\
x_2 & y_2 & 1 \\ 
\vdots & \vdots & \vdots\\
x_n & y_n & 1
\end{pmatrix}
\end{equation}
\noindent The constraint matrix is given by:
\begin{equation}
\mathbf{C} = 
\begin{pmatrix}
C_1 & 0 \\
0   & 0
\end{pmatrix},
\qquad
\text{where}
\qquad
C_{1} =
\begin{pmatrix}
0 & 0 & 2 \\
0 & -1 & 0 \\
2 & 0 & 0
\end{pmatrix}
\end{equation}
The scatter matrix is defined by:
\begin{equation}
\mathbf{S} = 
\begin{pmatrix}
    \mathbf{S}_1 & \mathbf{S}_2 \\
    \mathbf{S}_2^\mathrm{T} & \mathbf{S}_3
\end{pmatrix},
\quad \text{where} \quad
\left\{
\begin{aligned}
    \mathbf{S}_1 &= \mathbf{D}_1^\mathrm{T}\mathbf{D}_1 \\
    \mathbf{S}_2 &= \mathbf{D}_1^\mathrm{T}\mathbf{D}_2 \\
    \mathbf{S}_3 &= \mathbf{D}_2^\mathrm{T}\mathbf{D}_2
\end{aligned}
\right.
\end{equation}
The coefficient vector $\mathbf{a}$ is given by:
\begin{equation}
\mathbf{a} = 
\begin{pmatrix}
    \mathbf{a}_1 \\
    \mathbf{a}_2
\end{pmatrix},
\quad \text{where} \quad
\mathbf{a}_1 = 
\begin{pmatrix}
    a \\
    b \\
    c
\end{pmatrix}
\quad \text{and} \quad
\mathbf{a}_2 =
\begin{pmatrix}
    d \\
    e \\
    f
\end{pmatrix}
\end{equation}
Substituting the expressions for $\mathbf{S}$ and $\mathbf{a}$ into Eq.~\ref{eq-constraint}, we obtain the coupled vector equations:
\begin{align}\label{eqtwo-vector-1}
\mathbf{S}_1 \mathbf{a}_1 + \mathbf{S}_2 \mathbf{a}_2 &= \lambda\, \mathbf{C}_1 \mathbf{a}_1
\end{align}
\begin{align}\label{eqtwo-vector-2}
     \mathbf{S}_2^\mathrm{T} \mathbf{a}_1 + \mathbf{S}_3 \mathbf{a}_2 &= 0
\end{align}
From Equation \ref{eqtwo-vector-2}, $\mathbf{a_2}$ can be expressed as:
\begin{equation}
\mathbf{a}_2 = -\mathbf{S}_3^{-1} \mathbf{S}_2^\mathrm{T} \mathbf{a}_1
\end{equation}
After inserting the expression for $\mathbf{a_2}$ into Eq.~\ref{eqtwo-vector-1} and given that the matrix $\mathbf{C_1}$ is invertible, we obtain the following equation:
\begin{equation}
\mathbf{C}_1^{-1} \left( \mathbf{S}_1 - \mathbf{S}_2 \mathbf{S}_3^{-1} \mathbf{S}_2^\mathrm{T} \right) \mathbf{a}_1 = \lambda \mathbf{a}_1,
\end{equation}
where $\mathbf{a_1}$ needs to satisfy the condition:
\begin{equation}
\mathbf{a_1}^\mathrm{T} \mathbf{C_1} \mathbf{a_1} = 1    
\end{equation}
Thus, the fitting procedure reduces to solving the standard eigenvalue problem for the matrix  
\[
\mathrm{M} = \mathbf{C}_1^{-1} \left( \mathbf{S}_1 - \mathbf{S}_2 \mathbf{S}_3^{-1} \mathbf{S}_2^\mathrm{T} \right).
\]  
All eigenvalues $\lambda$ and their corresponding eigenvectors $\mathbf{a}_1$ of $\mathrm{M}$ are computed. Each eigenvector $\mathbf{a}_1$ represents a possible set of quadratic coefficients $(a,b,c)$ of the conic. The relevant solution is the eigenvector that satisfies the ellipse condition $ac - b^2 = 1$. The remaining linear and constant terms $(d, e, f)$ of the conic are then obtained from  
\[
\mathbf{a}_2 = -\mathbf{S}_3^{-1} \mathbf{S}_2^\mathrm{T} \mathbf{a}_1,
\]  
yielding the full parameter vector $\mathbf{a} = [\mathbf{a}_1; \mathbf{a}_2]$. The center of the fitted ellipse is defined as:

\begin{equation}
\begin{cases}
x_0 = \dfrac{2cd - 2be}{\,b^2 - ac\,}, \\[8pt]
y_0 = \dfrac{2ae - 2bd}{\,b^2 - ac\,},
\end{cases}
\qquad \text{with } b^2 - ac < 0.
\end{equation}
The minor and major axes of the ellipse are given by:

\begin{equation}
\begin{aligned}
a' &= \sqrt{\dfrac{2 \,\bigl(a f^{2} + c d^{2} + g b^{2} - 2 b d f - a c g \bigr)}{(b^{2} - ac)\,\bigl(\sqrt{(a-c)^{2} + 4b^{2}} - (a+c)\bigr)}}, \\[8pt]
b' &= \sqrt{\dfrac{2 \,\bigl(a f^{2} + c d^{2} + g b^{2} - 2 b d f - a c g \bigr)}{(b^{2} - ac)\,\bigl(-\sqrt{(a-c)^{2} + 4b^{2}} - (a+c)\bigr)}}, \\[8pt]
\end{aligned}
\end{equation}

\noindent The orientation angle is:
\begin{equation}
\phi \;=\;
\begin{cases}
0, & b=0 \ \text{and}\ a<c,\\[6pt]
\dfrac{\pi}{2}, & b=0 \ \text{and}\ a>c,\\[10pt]
\dfrac{1}{2}\arctan\!\Bigl(\dfrac{2b}{a-c}\Bigr), & b\neq 0 \ \text{and}\ a<c,\\[12pt]
\dfrac{1}{2}\arctan\!\Bigl(\dfrac{2b}{a-c}\Bigr)+\dfrac{\pi}{2}, & b\neq 0 \ \text{and}\ a>c,
\end{cases}
\qquad \phi \in [0,\pi).
\end{equation}
The parametric representation of the ellipse is given by:
\begin{equation}
\begin{cases}
x(t) = x_0 + a' \cos t \cos\phi - b' \sin t \sin\phi, \\[8pt]
y(t) = y_0 + a' \cos t \sin\phi + b' \sin t \cos\phi,
\end{cases}
\qquad t \in [0,2\pi).
\end{equation}
The observed \HI\ diameter is defined as:  
\begin{equation}
D_{\mathrm{HI,obs}} = 2a',    
\end{equation}
where \(a'\) is the semimajor axis of the fitted ellipse. 

\section{Surface-density noise map}
After integrating $N_{\rm ch}(x,y)$ independent spectral channels, the flux uncertainty at each pixel is given by
\begin{equation}\label{eq:flux_noise}
    \sigma_{\rm flux}(x,y) = \sigma_{\rm ch} \, \Delta v \, \sqrt{N_{\rm ch}(x,y)}
    \quad [\mathrm{Jy\,beam^{-1}\,km\,s^{-1}}],
\end{equation}
where $\sigma_{\rm ch}$ is the median per-channel noise and $\Delta v$ is the
channel width.

To convert this flux uncertainty to \HI\ column density, we use the standard formula:
\begin{equation}\label{eq:NHI_Tb}
    N_{\rm HI} = 1.823 \times 10^{18} \int T_{\rm b} \, dv 
    \quad [\mathrm{cm^{-2}}],
\end{equation}
where $T_{\rm b}$ is the brightness temperature in Kelvin and the integral is
in $\mathrm{km\,s^{-1}}$. For a Gaussian synthesized beam at 1.42\,GHz, the
conversion between flux density and brightness temperature is
\begin{equation}\label{eq:Tb_flux}
    T_{\rm b} = \frac{c^{2}}{2 k_{\rm B} \nu^{2}} \frac{S_{\nu}}{\Omega_{\rm beam}}\,,
\end{equation}
where $S_{\nu}$ is the flux density, $\Omega_{\rm beam} = (\pi / 4\ln 2)\,
B_{\rm maj}\,B_{\rm min}$ is the beam solid angle in $\mathrm{arcsec^{2}}$,
and $B_{\rm maj}$ and $B_{\rm min}$ are the beam FWHM in arcsec. Substituting
Eq.~\ref{eq:Tb_flux} into Eq.~\ref{eq:NHI_Tb} gives
\begin{equation}\label{eq:NHI_flux}
    N_{\rm HI} = \frac{1.104 \times 10^{21}}{\Omega_{\rm beam}}
    \, S_{\nu} \, \Delta v
    \quad [\mathrm{cm^{-2}}],
\end{equation}
where $S_{\nu}$ is in $\mathrm{Jy\,beam^{-1}}$ and $\Delta v$ in
$\mathrm{km\,s^{-1}}$.

Finally, we convert column density to mass surface density:
\begin{equation}\label{eq:Sigma_NHI}
    \Sigma_{\rm HI} = N_{\rm HI} \times m_{\rm H} 
    \times \left( \frac{1\,\mathrm{pc}}{1\,\mathrm{cm}} \right)^{2} 
    \times \frac{1}{M_{\odot}}
    = \frac{N_{\rm HI}}{1.248 \times 10^{20}}
    \quad [M_{\odot}\,\mathrm{pc^{-2}}],
\end{equation}
where $m_{\rm H} = 1.674 \times 10^{-24}\,\mathrm{g}$ is the hydrogen atom mass,
$1\,\mathrm{pc} = 3.086 \times 10^{18}\,\mathrm{cm}$, and
$M_{\odot} = 1.989 \times 10^{33}\,\mathrm{g}$.

Combining Equations~\ref{eq:flux_noise}, \ref{eq:NHI_flux}, and \ref{eq:Sigma_NHI},
the surface-density uncertainty at each pixel is
\begin{equation}\label{eq:sigma_Sigma}
    \sigma_{\Sigma}(x,y) = 
    \frac{C_{\rm conv} \, \sigma_{\rm ch} \, \Delta v \, \sqrt{N_{\rm ch}(x,y)}}
    {\Omega_{\rm beam}}\,,
\end{equation}
where $C_{\rm conv} = 1.104 \times 10^{21} / 1.248 \times 10^{20}$ is the
combined unit-conversion factor from integrated flux
($\mathrm{Jy\,beam^{-1}\,km\,s^{-1}}$) to mass surface density
($M_{\odot}\,\mathrm{pc^{-2}}$), and $\Omega_{\rm beam} = (\pi / 4\ln 2)\,
B_{\rm maj}\,B_{\rm min}$ is the beam solid angle in $\mathrm{arcsec^{2}}$.

\section{HI nondetected HCG members (beam-size upper limits)}\label{app:upperlimits}
Table~\ref{table:upperlimits} lists the \HI\ nondetected HCG members. We use their major 
axis beam-size as upper limits to their \HI\ diameter (see Sect.~\ref{results}). 
For each of them, the upper limit on $D_{\rm HI}$ is the major-axis beam $B_{\rm maj}$ converted to a linear 
size at the group distance, and $D_{25}$ is taken from HyperLeda. These galaxies are treated as 
left-censored data, and we use the Kaplan--Meier and Gehan statistics to analyze them.
{\small
\begin{table}[!htbp]
\centering
\caption{\label{table:upperlimits}HCG members entering the residual statistics as beam-size upper limits.}
\scriptsize
\setlength{\tabcolsep}{4pt}
\begin{tabular}{llccccc}
\toprule \toprule
HCG & member & phase & $B_{\rm maj}$ & $D$ & $D_{25}$ & $D_{\rm HI}$ \\
    &        &       & [\arcsec] & [Mpc] & [kpc] & [kpc] \\
\midrule
7 & HCG7b & 1 & 35.0 & 50 & 18.3 & $<8.5$ \\
10 & HCG10b & 1 & 61.6 & 52 & 33.9 & $<15.5$ \\
10 & HCG10c & 1 & 61.6 & 49 & 25.3 & $<14.6$ \\
15 & HCG15a & 3c & 60.7 & 95 & 35.6 & $<28.0$ \\
15 & HCG15b & 3c & 60.7 & 97 & 26.9 & $<28.5$ \\
15 & HCG15c & 3c & 60.7 & 98 & 24.3 & $<28.8$ \\
15 & HCG15d & 3c & 60.7 & 84 & 22.3 & $<24.7$ \\
15 & HCG15e & 3c & 60.7 & 97 & 20.0 & $<28.5$ \\
15 & HCG15f$^{a}$ & 3c & 60.7 & 84 & 16.5 & $<24.7$ \\
19 & HCG19a & 1 & 41.0 & 53 & 18.5 & $<10.5$ \\
22 & HCG22a & 3c & 50.9 & 38 & 28.4 & $<9.4$ \\
22 & HCG22b & 3c & 50.9 & 37 & 8.2 & $<9.1$ \\
22 & NGC1188 & 3c & 50.9 & 37 & 12.9 & $<9.1$ \\
23 & HCG23c & 1 & 25.3 & 67 & 17.4 & $<8.2$ \\
25 & HCG25d & 1 & 65.4 & 83 & 9.8 & $<26.3$ \\
25 & HCG25f & 1 & 65.4 & 81 & 10.1 & $<25.7$ \\
26 & HCG26b & 1 & 25.5 & 122 & 27.5 & $<15.1$ \\
26 & HCG26c & 1 & 25.5 & 130 & 14.7 & $<16.1$ \\
26 & HCG26d & 1 & 25.5 & 118 & 22.7 & $<14.6$ \\
26 & HCG26f & 1 & 25.5 & 131 & 8.3 & $<16.2$ \\
26 & HCG26g & 1 & 25.5 & 121 & 8.3 & $<15.0$ \\
30 & HCG30a & 3a & 57.9 & 62 & 26.1 & $<17.4$ \\
30 & HCG30b & 3a & 57.9 & 61 & 21.3 & $<17.1$ \\
30 & HCG30c & 3a & 57.9 & 59 & 10.6 & $<16.5$ \\
30 & HCG30d & 3a & 57.9 & 61 & 10.0 & $<17.1$ \\
33 & HCG33a & 3c & 17.5 & 104 & 27.6 & $<8.8$ \\
33 & HCG33b & 3c & 17.5 & 110 & 32.7 & $<9.3$ \\
33 & HCG33d & 3c & 17.5 & 106 & 17.7 & $<9.0$ \\
37 & HCG37a & 3a & 49.5 & 97 & 56.3 & $<23.3$ \\
37 & HCG37b & 3a & 49.5 & 97 & 40.8 & $<23.3$ \\
37 & HCG37c & 3a & 49.5 & 108 & 13.7 & $<25.9$ \\
37 & HCG37d & 3a & 49.5 & 89 & 9.8 & $<21.4$ \\
37 & HCG37e & 3a & 49.5 & 91 & 9.6 & $<21.9$ \\
40 & HCG40a & 2 & 58.8 & 99 & 38.0 & $<28.2$ \\
40 & HCG40b & 2 & 58.8 & 102 & 26.4 & $<29.1$ \\
40 & HCG40d$^{a}$ & 2 & 58.8 & 98 & 26.0 & $<27.9$ \\
40 & HCG40e & 2 & 58.8 & 99 & 19.5 & $<28.2$ \\
56 & HCG56b & 3c & 23.5 & 113 & 24.9 & $<12.9$ \\
56 & HCG56c & 3c & 23.5 & 116 & 27.4 & $<13.2$ \\
56 & HCG56d & 3c & 23.5 & 120 & 16.7 & $<13.7$ \\
56 & HCG56e & 3c & 23.5 & 113 & 14.3 & $<12.9$ \\
58 & HCG58e$^{a}$ & 2 & 65.8 & 83 & 16.7 & $<26.5$ \\
62 & HCG62a & 3a & 24.8 & 62 & 48.5 & $<7.5$ \\
62 & HCG62b & 3a & 24.8 & 53 & 30.1 & $<6.4$ \\
62 & HCG62c & 3a & 24.8 & 62 & 27.9 & $<7.5$ \\
62 & HCG62d & 3a & 24.8 & 59 & 7.8 & $<7.1$ \\
68 & HCG68a & 3c & 58.3 & 32 & 22.3 & $<9.0$ \\
68 & HCG68b & 3c & 58.3 & 40 & 35.1 & $<11.3$ \\
68 & HCG68d & 3c & 58.3 & 35 & 11.7 & $<9.9$ \\
68 & HCG68e & 3c & 58.3 & 35 & 11.4 & $<9.9$ \\
71 & HCG71b & 2 & 26.0 & 133 & 24.4 & $<16.8$ \\
79 & HCG79a & 2 & 20.8 & 58 & 30.0 & $<5.8$ \\
79 & HCG79b & 2 & 20.8 & 60 & 40.0 & $<6.0$ \\
79 & HCG79c & 2 & 20.8 & 56 & 26.4 & $<5.6$ \\
90 & HCG90b & 3c & 49.1 & 31 & 10.4 & $<7.4$ \\
90 & HCG90c & 3c & 49.1 & 35 & 19.0 & $<8.3$ \\
90 & HCG90d & 3c & 49.1 & 36 & 26.3 & $<8.6$ \\
91 & HCG91d & 2 & 51.3 & 92 & 13.7 & $<22.9$ \\
92 & HCG92b & 3a & 15.0 & 73 & 50.9 & $<5.3$ \\
92 & HCG92c & 3a & 15.0 & 90 & 43.4 & $<6.5$ \\
92 & HCG92d & 3a & 15.0 & 87 & 37.4 & $<6.3$ \\
92 & HCG92e & 3a & 15.0 & 87 & 20.1 & $<6.3$ \\
93 & HCG93a & 3c & 59.7 & 64 & 26.9 & $<18.5$ \\
93 & HCG93c & 3c & 59.7 & 64 & 19.5 & $<18.5$ \\
93 & HCG93d & 3c & 59.7 & 64 & 12.9 & $<18.5$ \\
96 & HCG96b & 2 & 26.9 & 115 & 25.4 & $<15.0$ \\
96 & HCG96c & 2 & 26.9 & 117 & 11.8 & $<15.3$ \\
96 & HCG96d & 2 & 26.9 & 120 & 8.2 & $<15.7$ \\
97 & HCG97a & 3c & 63.2 & 91 & 42.0 & $<27.9$ \\
97 & HCG97b & 3c & 63.2 & 92 & 32.9 & $<28.2$ \\
97 & HCG97c & 3c & 63.2 & 76 & 21.6 & $<23.3$ \\
97 & HCG97d & 3c & 63.2 & 80 & 39.5 & $<24.5$ \\
97 & HCG97e & 3c & 63.2 & 85 & 11.0 & $<26.0$ \\
\bottomrule
\end{tabular}
\tablefoot{The table lists the \HI\ non-detections plus the members marked $^{a}$, which are detected in \HI\ but spatially unresolved ($D_{\rm HI}<$ beam). $D_{\rm HI}$ is the major-axis beam ($B_{\rm maj}$) converted to kpc at the group distance; $D_{25}$ is from HyperLeda.}
\end{table}
}

\section{Reproducibility}\label{reproducibility}

Alongside the scientific results, we have built the analysis so that a single
command can reproduce the full paper, from the input maps to the final
manuscript. The reproducible chain begins with the \HI\ moment-zero maps. The
raw data cubes and source masks used to create these maps were
produced and released by \citet{2023A&A...670A..21J}. Following the FAIR principles, we provide the code and inputs required
to reproduce the paper in a publicly available GitHub repository (\url{https://github.com/ianjarog/galaxydisksize}).
\\ \indent We have split the framework into  a reusable Python library and
a \texttt{Snakemake} workflow that executes the complete analysis. The
scientific routines (ellipse fitting, baseline and size--mass relations, 
surface-density conversions, and censored-data statistics) are
implemented in the installable Python package
\texttt{galaxydisksize}. This package can be used independently for other
samples and analyses. The \texttt{Snakemake} workflow connects these routines
into an end-to-end directed acyclic graph, where each output file is generated
by a single rule. This design ensures that each output has a single origin 
and prevents multiple analysis steps from accidentally overwriting the same figure or table.
\\ \indent We organize the workflow into two tiers. The first, the catalog tier, performs an ellipse fitting 
to the moment-zero maps to measure the \HI\ diameter and mass of each galaxy, 
and tabulates them together with the optical diameters. 
This is the computationally expensive stage because it requires the source masks used to produce the moment maps. 
These masks are available at \url{https://zenodo.org/records/6909872}. So, this step is rerun only when the input maps change. 
The second, the statistics tier, uses the per-galaxy measurements in the first step as input to fit the scaling relations, 
perform the residual and survival analyses, and generate every figure, table, and the manuscript. 
\\ \indent We provide a pinned \texttt{conda} environment file together with
container recipes to allow users to easily recreate the software environment used
for the analysis. We also provide a browser-executable configuration that
allows a subset of the results to be reproduced without a local installation.

\end{appendix}
\end{document}